\documentclass{aa}  

\usepackage{graphicx}
\usepackage{multirow}
\usepackage{subfigure}
\usepackage{textcomp}
\usepackage{xcolor}
\usepackage{booktabs}
\usepackage{tablefootnote}
\usepackage{footmisc}
\usepackage{tabularx}
\usepackage{float}
\usepackage{txfonts}
\usepackage[colorlinks=true, allcolors=blue]{hyperref}

\renewcommand{\textbf}[1]{#1}

\begin{document}

   \title{Substellar population of the young massive cluster RCW \,36 in Vela\thanks{Based on observations collected at the European Southern Observatory under ESO program 0104.C-0369}}
   \titlerunning{Substellar population in RCW\,36}
   
   \author{A.~R.~G.~do Brito do Vale\inst{\ref{IA},\ref{ia2}, \ref{LAB}}, 
   K. Mu\v{z}i\'c\inst{\ref{IA}, \ref{ia2}}, 
   H. Bouy\inst{\ref{LAB}, \ref{herve}},
   V. Almendros-Abad\inst{\ref{victor}},
   A. Bayo\inst{\ref{amelia}},
   D. Capela\inst{\ref{daniel}},
   A. Scholz\inst{\ref{daniel}},
   A. Bik\inst{\ref{bik}},
   G. Su\'arez\inst{\ref{genaro_NY}},
   L. Cieza\inst{\ref{Chile1}, \ref{Chile2}}
   K. Pe\~na Ramírez\inst{\ref{karla}},
   E.~Bertin\inst{\ref{inst_ups}, \ref{inst_cfht}},
   R. Schödel\inst{\ref{rainer}}
          }
   \authorrunning{A.~M.~R.~G.~do Brito do Vale}
   \institute{Instituto de Astrofĩsica e Ciências do Espaço (IA), Faculdade de Ci\^{e}ncias, Universidade de Lisboa, Ed. C8, Campo Grande, P-1749-016 Lisboa, Portugal\label{IA}\\
              \email{fc47932@alunos.fc.ul.pt}
        \and Departamento de Física, Faculdade de Ciências, Universidade de Lisboa, Edifício C8, Campo Grande, 1749-016 Lisbon, Portugal\label{ia2}
        \and Laboratoire d'Astrophysique de Bordeaux (LAB), Université de Bordeaux, Bât. B18N, Allée Geoffroy Saint-Hilaire CS 50023, 33615 PESSAC CEDEX, France\label{LAB}
        \and Institut universitaire de France (IUF), 1 rue Descartes, 75231 Paris CEDEX 05 \label{herve}
        \and NSF NOIRLab/Vera C. Rubin Observatory, Casilla 603, La Serena, Chile\label{karla}
        \and Istituto Nazionale di Astrofisica (INAF) – Osservatorio Astronomico di Palermo, Piazza del Parlamento 1, 90134 Palermo, Italy \label{victor}
        \and European Southern Observatory, Karl-Schwarzschild-Strasse 2, 85748 Garching bei M\"unchen, Germany \label{amelia}
        \and SUPA, School of Physics \& Astronomy, University of St Andrews, North Haugh, St Andrews, KY16 9SS, United Kingdom\label{daniel}
        \and Department of Astronomy, Stockholm University, AlbaNova University Center, 10691 Stockholm, Sweden\label{bik}
        \and Department of Astrophysics, American Museum of Natural History, Central Park West at 79th Street, NY 10024, USA\label{genaro_NY}
        \and Instituto de Estudios Astrofisicos, Facultad de Ingenieriıa y Ciencias, Universidad Diego Portales, Av. Ejercito 441, Santiago, Chile\label{Chile1}
        \and Millennium Nucleus on Young Exoplanets and their Moons (YEMS), Chile \label{Chile2}
        \and  Universit\'e Paris-Saclay, Universit\'e Paris Cit\'e, CEA, CNRS, AIM, F-91191, Gif-sur-Yvette, France\label{inst_ups}
        \and Canada-France-Hawaii Telescope Corporation, 65-1238 Mamalahoa Highway, Kamuela, HI96743, USA\label{inst_cfht}
        \and Instituto de Astrofísica de Andalucía (CSIC), Glorieta de la Astronomía s/n, 18008 Granada, Spain\label{rainer}
        }

   \date{Received; accepted}

 
  \abstract
   {The shape of the initial mass function (IMF) remains a fundamental yet contentious topic in the study of stellar formation and evolution. It is imperative to understand the potential variability of the IMF across different young regions. This study examines the IMF within the young massive cluster RCW\,36 situated in the Vela Molecular Ridge, comparable with the Orion Nebula Cluster in terms of stellar surface density.}
   {The primary objective of this research is to construct the most comprehensive census of the stellar population in RCW\,36 to date and determine the first ever IMF and star to brown dwarf (BD) ratio for the cluster.}
   {We used state-of-the art observational techniques, drawing on new GLAO observations conducted with HAWK-I/VLT in addition to archival data from 2MASS, SOFI/NTT, and new kinematics from Gaia\,DR3. To enhance photometric accuracy and source extraction, we employed \textsc{DeNeb}, an advanced deep learning algorithm capable of removing the complex filamentary nebula in our images. Statistical comparisons of color-magnitude diagrams were performed between RCW\,36 and a control field, also obtained using HAWK-I under the same mode, to assign membership weights for the sources in our field. Mass estimates to individual sources were also derived through comparison with model isochrones in order to determine the IMF using the membership weights.}
   {We found a new distance of $954\pm40\,$pc. We determined the IMF for RCW\,36 down to $\sim0.03 M_{\odot}$, characterized by a broken power law ($dN/dM\propto M^{-\alpha}$) with $\alpha = 1.62 \pm 0.03$ ($0.20M_{\odot}-20M_{\odot}$) and $\alpha=0.46\pm0.14$ ($0.03M_{\odot}-0.20M_{\odot}$). We also determined the star-BD ratio to be $2-5$, in agreement with other Galactic clusters. Lastly, through a study of the differences in the IMF within and outside $0.2\,$pc and the cumulative mass distributions for low-mass and intermediate to high-mass sources, we also detected signs of possible mass segregation within RCW\,36, which should be primordial.}
  {RCW\,36 shares many characteristics with other young massive clusters, such as a shallower than Salpeter high-mass slope and the possibility of mass segregation. The flatter lower-mass regime of the IMF is similar to most Galactic clusters. The star-BD ratio is also in line with the observed values in other clusters, independent of their inherent properties.}  
\maketitle
%

\section{Introduction}
Embedded stellar clusters are regions of active, or very recent, star formation in the Milky Way that arise in the interior of cold, dense molecular gas clouds. In these clouds, a broad range of stellar and substellar cores formed from local gravitational instabilities eventually evolve through accretion and contraction processes to shed the gas and dust around them and become detectable first in the near infrared (NIR) and eventually in the optical \citep{Ladas2003, Luhman2012, bate12}. These stellar populations are of great interest in the subject of star formation because they comprise a direct unaltered consequence of the star formation processes before kinematics and stellar evolution can alter the properties of the cluster. One very important property of such young clusters is their initial distribution of masses, the initial mass function (IMF).\\
\begin{figure*}[t]
   \centering
   \includegraphics[width=1.0\textwidth]{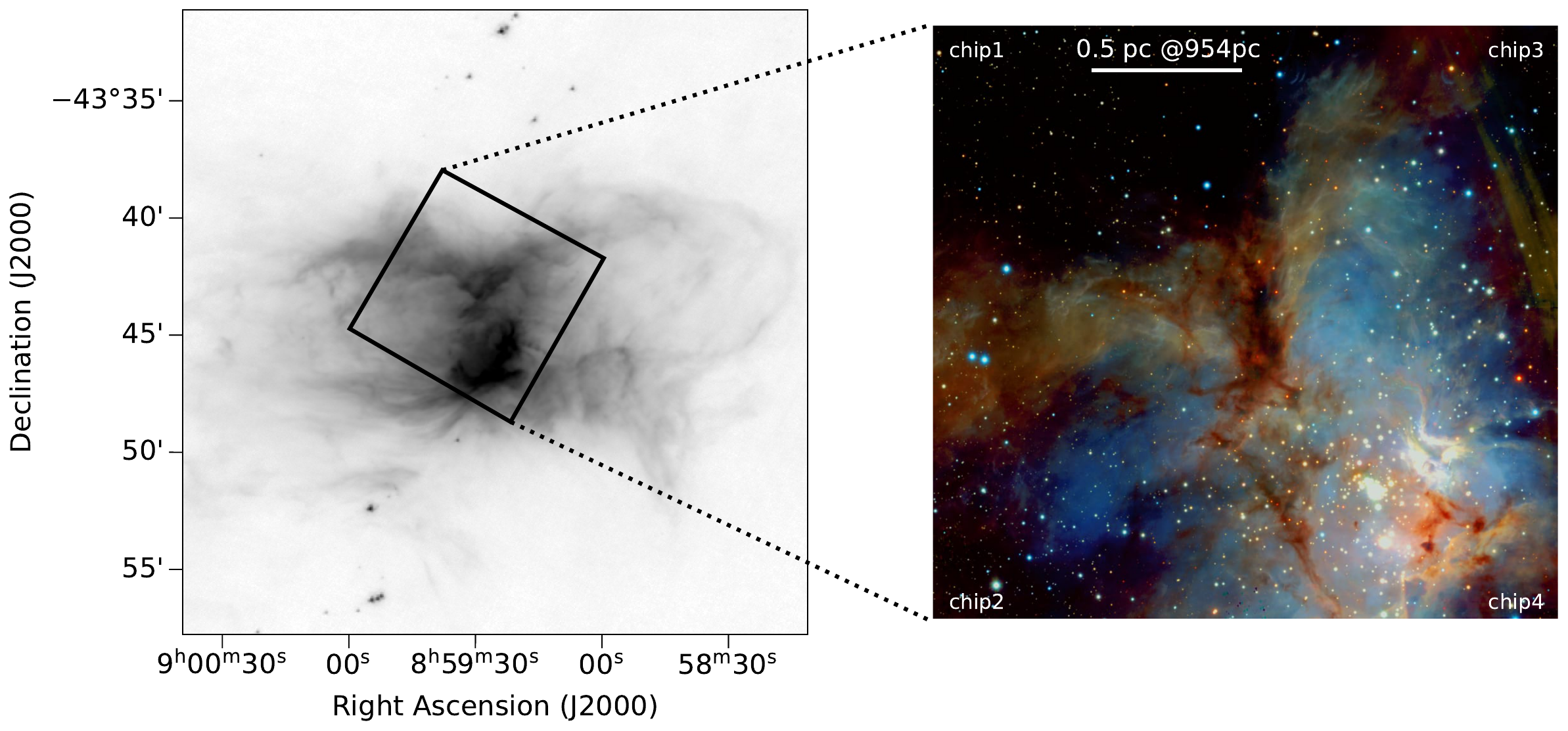}
      \caption{Left: $Herschel$ PACS \textbf{70$\mu$m} image of the region in Vela associated with the cluster RCW\,36. The black rectangle marks the area under study. Right: $JHK_s$ bands  colour mosaic obtained with HAWK-I/VLT. The position of the four detectors is marked.}
         \label{fig:onsky}
\end{figure*}
Initially, \cite{Salpeter1955} observed that the IMF of the solar neighbourhood for sources over $0.5M_{\odot}$ could be described by a power law, $dN/dM \propto M^{-\alpha}$, with the slope $\alpha=2.35$, which would later become known as the Salpeter slope, and kickstarted the debate on the possible universality of the shape of the IMF across different star forming environments. In the 70 years that followed this discovery, studies of young clusters in the solar neighbourhood and the Galactic disk have shown that the Salpeter slope is only a good approximation of the IMF down to a stellar mass of $~0.5M_{\odot}$. For lower masses, this slope has been observed in these same clusters and associations to be significantly flatter ($\alpha\sim0-1$, \citealt{Muench+2000, Briceno+2002, Adams+2002,DaRio2012, Scholz+2013,AlvesdeOliveira2012,Alves2013,muzic15, Bouy+2015}; see Figure 1 in the review by \citealt{Hennebelle+2024}). In fact, currently, it is most common to describe the IMF as a set of different power laws for different stellar mass regimes \citep{Kroupa2001} or a log-normal distribution \citep{Chabrier+2005} both of which accommodate the observed flattening of the lower-mass regime.\\

This flattening of the $\alpha$ that has been observed in both the more sparsely distributed and less dense clusters and associations in Chameleon \citep{muzic15, Kubiak+2021}, Lupus \citep{muzic15, Galli+2020}, Perseus \citep{scholz12a, Scholz+2013, Alves2013}, the Taurus Molecular Clouds \citep{Briceno+2002, Luhman2018}, Corona Australis \citep{Muzic+2025}, Rosette Nebula \citep{Muzic+2019,Almendros-Abad2023}, and in the more massive clusters in the Orion Nebula \citep{Muench2002, DaRio2012}, Carina Nebula \citep{Hur+2012,Lim2013,Andersen2017, Rom+2025} and the Vela Molecular Ridge (VMR) \citep{Massi+2006,Muzic2017} does not necessarily imply a universal IMF. It does, however, clearly show that the IMF shows remarkable regularity in its shape and characteristics across different clusters and associations within our galaxy \citep{Hennebelle+2024}, regardless of their inherent characteristics.\\
\cite{Damian+2021} analysed the IMF of eight clusters with ages $<5$\,Myr and varying Galactocentric distances ($~6-12$\,kpc) and report no strong evidence for an environmental effect in the underlying form of the IMF. The same conclusion can also be found in \cite{Offner2014}. However, observational inconsistencies in the $\alpha$ obtained in the same mass range in clusters with different properties (metallicity, \citealt{Bate+2025,Li+2023}; stellar surface density, \citealt{Briceno+2002, Bate+2005}; presence of OB stars \citealt{Whitworth_Zinnecker2004,padoan2004, Luhman2012}) are still unquestionably observed, and a real universal shape of the IMF is not agreed upon. \\
More examples can be found in the work of \cite{Dib2014}, who found evidence against the universality of the IMF in a Bayesian analysis on a sample of 8 young Galactic clusters, and in the shallower-than-Salpeter \textbf{high-mass} $\alpha$ that has been observed in young massive clusters (YMCs) such as RCW\,38 \citep{Muzic2017}, Collinder\,69 \citep{Bayo+2011}, Westerlund\,I \citep{Lim2013,Andersen2017}, and the Arches cluster towards the \textbf{Galactic} centre \citep{Hosek2019, Habibi2013}, which differs from the classical view of the universality of the Salpeter slope in the \textbf{high-mass} regime of the IMF.\\

In this paper, we present new ground layer adaptive optics (GLAO) observations of the young massive cluster RCW\,36 \citep{Rodgers1960, Pettersson2008}, located along a filamentary structure in cloud C of the VMR \citep{Murphy91}, which has been found to be associated with many features related to ongoing intermediate to \textbf{high-mass} star formation (HII region, \citealt{Gum1955, Anderson+2014}; $H_2O$ maser, \citealt{BrazScalise1982}; $\gamma$-rays, \citealt{Peron+2024,Peron2025}). Previous studies of this cluster have been able to identify a rich population of massive and intermediate-mass stars \citep{Baba04} located at a distance of $\sim700$pc \citep{Liseau92}, including a pair of late O stars \citep{ellerbroek13}. The high surface density of $10^3$-$10^4$ stars per squared parsec in its central region \citep{Baba04,Kuhn+2015III}, which is comparable with the ONC \citep{Baba04, Muzic+2019}, and its young age of $\leq1.1$\,Myr \citep{ellerbroek13, Kuhn+2015III, Sano2018} make RCW\,36 a great target to study the effects of environments significantly different than the nearby SFRs where the IMF has been best studied. Previous studies of this cluster have been focused on identifying earlier type stars (earlier than $~$K, \citealt{Baba04,ellerbroek13}), but the lowest mass content has not yet been thoroughly explored.\\
RCW\,36 has also been observed in the scope of the Massive Young star-forming Complex Study in Infrared and X-rays (MYStIX) project \citep{broos13,kuhn14, Kuhn+2015III}, which was able to determine 384 candidate members from infrared and X-ray data.
Our new GLAO observations together with a novel deep learning algorithm able to remove the nebula component in our images provide the deepest look yet into the embedded stellar population of RCW\,36, deriving the first ever IMF for RCW\,36, complete down to $\sim0.03M_{\odot}-0.05M_{\odot}$.\\

This paper is organized as follows. In Section \ref{sec:observations_and_data_reduction}, we present the observations, data reduction, and further processing done. In Section \ref{sec:photometry} we present the photometry, discuss the differences due to the advanced processing described in the former section, and showcase the properties of our final $JHK_s$ catalogue. In Section \ref{s:gaia_distance}, we present our new distance estimation to RCW\,36 using the Gaia DR3 \citep{Gaia_DR3} and the previous known members. Then, in Section \ref{s:membership} we describe the statistical procedure used to assign membership \textbf{weights} to the sources in our catalogue. In Section \ref{ss:masses_ext}, we describe the process to obtain the mass probability distribution functions (PDFs) for all sources in our catalogue. In Section \ref{s:imf} we employ the mass PDFs and the membership weights to determine the IMF and then in Section \ref{ss:star_bds_ratio} the star to brown dwarf ratio. Finally, we analyse the possibility of mass segregation in RCW\,36 in Section \ref{ss:mass_segregation} and summarise our final conclusions in Section \ref{s:conclusions}.
\section{Observations and data processing}\label{sec:observations_and_data_reduction}
\subsection{HAWK-I imaging}
Observations of RCW\,36, as well as of another field outside the
cluster area (the control field) were performed using the High Acuity Wide-field $K_s$-band  Imager (HAWK-I; \citealt{pirard2004,kissler-patig2008}) at the Very Large Telescope.\footnote{ESO programme ID 0104.C-0369}
The instrument was used together with the adaptive optics system GRAAL \citep{paufique2010, Arsenault+2017}, with a four-laser-guide-star system and a deformable secondary mirror to correct for the ground layer atmospheric turbulence.
HAWK-I provides an on-sky field of view of $7.5'\times7.5'$, with a cross-shaped gap of $15''$ between its four detectors, and a pixel scale of $0.106''$.
The data were obtained in service mode using the $J$-, $H$-, and $K_s$-band filters. 
To estimate the amount of contamination by field stars, we observed a control field, located $\sim 1^\circ$ away from the centre of RCW\,36, along the same \textbf{Galactic} latitude. The coordinates of the cluster and the control field, along with other details of the observations, are given in Table~\ref{tab:obs}.\\
For both the cluster and the control field, a sequence of dithered images was acquired, using randomly selected offsets within a $30''$ box. In the case of the cluster, we also visited a nearby field ($\sim8.5'$ away) containing fewer stars and no extended component \textbf{(nebula)}, which serves to estimate the sky contribution. In the case of the control field, the sky has been determined from dithered science exposures.\\
\begin{table*}
\centering
\caption{Summary of the observations. \label{tab:obs}}
\begin{tabular}{lccccccc}\hline\hline
Field   & $\alpha$ (J2000) & $\delta$ (J2000) & date & filter & exp.time (s) & FWHM (\farcs) & airmass\\
\hline
RCW\,36 & 08:59:30 & -43:43:17 & 2020-01-31, 2020-03-17 & $J$ & 2520  & 0.55 & 1.1\\
RCW\,36 &  08:59:30 & -43:43:17  &  2020-02-06  & $H$ & 765 & 0.45 & 1.1 \\
RCW\,36 &  08:59:30 & -43:43:17  & 2020-02-06 & $Ks$ &  510  & 0.40 & 1.1 \\
control field & 09:03:00 & -44:27:38   & 2020-03-01 & $J$ & 810 & 0.50 & 1.1  \\
control field & 09:03:00 & -44:27:38   & 2020-03-01 & $H$ & 306 & 0.40 & 1.1 \\
control field & 09:03:00 & -44:27:38   & 2020-03-01 & $Ks$& 306 & 0.30 & 1.1 \\
\hline
\end{tabular}
\end{table*}
\begin{figure*}
    \centering
    \includegraphics[width=\textwidth]{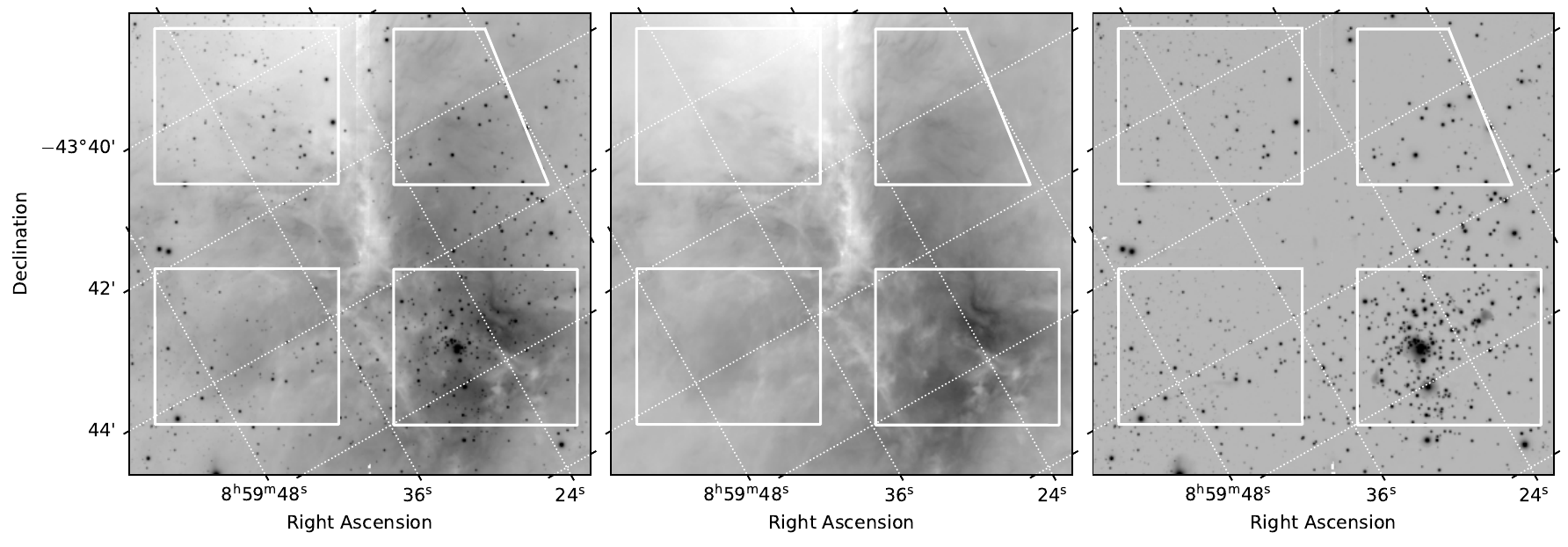}
    \caption{Left: Original HAWK-I J-band image of RCW\,36. Centre: HAWK-I J-band image with only the nebular component. Right: HAWK-I J-band image with only the point sources (hereby called denebulised). The white contours trace the regions used in the present analysis; the different contour in chip 3 is due to the artefact described in Section \ref{sec:observations_and_data_reduction}. The dotted lines refer to the coordinate grid lines of constant right ascension or declination.}
    \label{fig:rcw36_deneb}
\end{figure*}
Standard near-infrared data reduction, including dark subtraction, sky subtraction, flat-field, bad pixel correction, detector linearity correction, and creation of final mosaics was performed using ESO’s HAWK-I pipeline \textbf{for both the science field and the control field}.
The reduced $H-$ and $K_s-$ bands data of the science field (taken on the same night) show an arch-shaped artefact in the upper-right corner of the mosaic. It is likely caused by scattered light from the nearly full Moon which was located approximately 70$^\circ$ to the north-west of the target. The artefact was not present in other data taken that night, nor in previous $K_s$-band data of the same field taken about a week earlier (although the latter were of lower quality and therefore excluded). The region affected by the artefact was visually identified in the science images and ignored in the further analysis.
In Fig.~\ref{fig:onsky} we show the region covered by the HAWK-I observations, along with a $JHK_s$ colour composite image; the nomenclature associated. In this figure, the aforementioned artefact can be seen in the top right corner as the angled stripped shapes, which resulted in a removal of a part \textbf{of one of the HAWK-I detectors, chip 3,} from the analysis (\textbf{see Figure \ref{fig:onsky} for the detector identification and Figure \ref{fig:rcw36_deneb} for final areas kept in the analysis}).
\subsection{Processing}
\label{sec:DeNeb}
RCW\,36 is embedded in a thick, spatially variable nebula that is part of the larger VMR complex. The presence of this diffuse nebulosity complicates the identification of the point sources in RCW\,36 and worsens our capabilities of achieving accurate photometric measurements. To overcome this, we employ \textsc{DeNeb} (Bertin et al, in prep), a novel deep learning algorithm which is able to identify and separate the nebula and point source components in our NIR images. \textsc{DeNeb} employs a Convolutional Neural Network (CNN) model (Pytorch; \citealt{PyTorch2019}) to exploit subtle spatial correlations in the input data across various scales and directions and infer extended emissions from astronomical images, effectively suppressing contributions from other sources like stars, galaxies, and background noise. This is achieved by combining the outputs of millions of convolution kernels, that were automatically adapted through training on a specially curated training set of pure nebulae. Thus, after processing, a single band image of RCW\,36 will be split onto two new images: one featuring only the sources (both point sources and extended sources) and another featuring only the extended nebula.\\
In Fig.~\ref{fig:rcw36_deneb}, we present the original HAWK-I $J$-band image (left) and the two outputs from applying \textsc{DeNeb}: the nebular component (centre) and the point sources (right; hereby called denebulised images).
In Section \ref{sec:photometry} we introduce the photometry extracted from the original and denebulised $JHK_s$ HAWK-I data and discuss their differences. In addition, a more complete discussion showcasing the different populations in the $JHK_s$ catalogue of each set of images can also be found in Appendix~\ref{appendix:Deneb_analysis}.
\section{Photometry}
\label{sec:photometry}
\subsection{Extraction and calibration}
\begin{figure}
    \centering
    \includegraphics[width=0.5\textwidth]{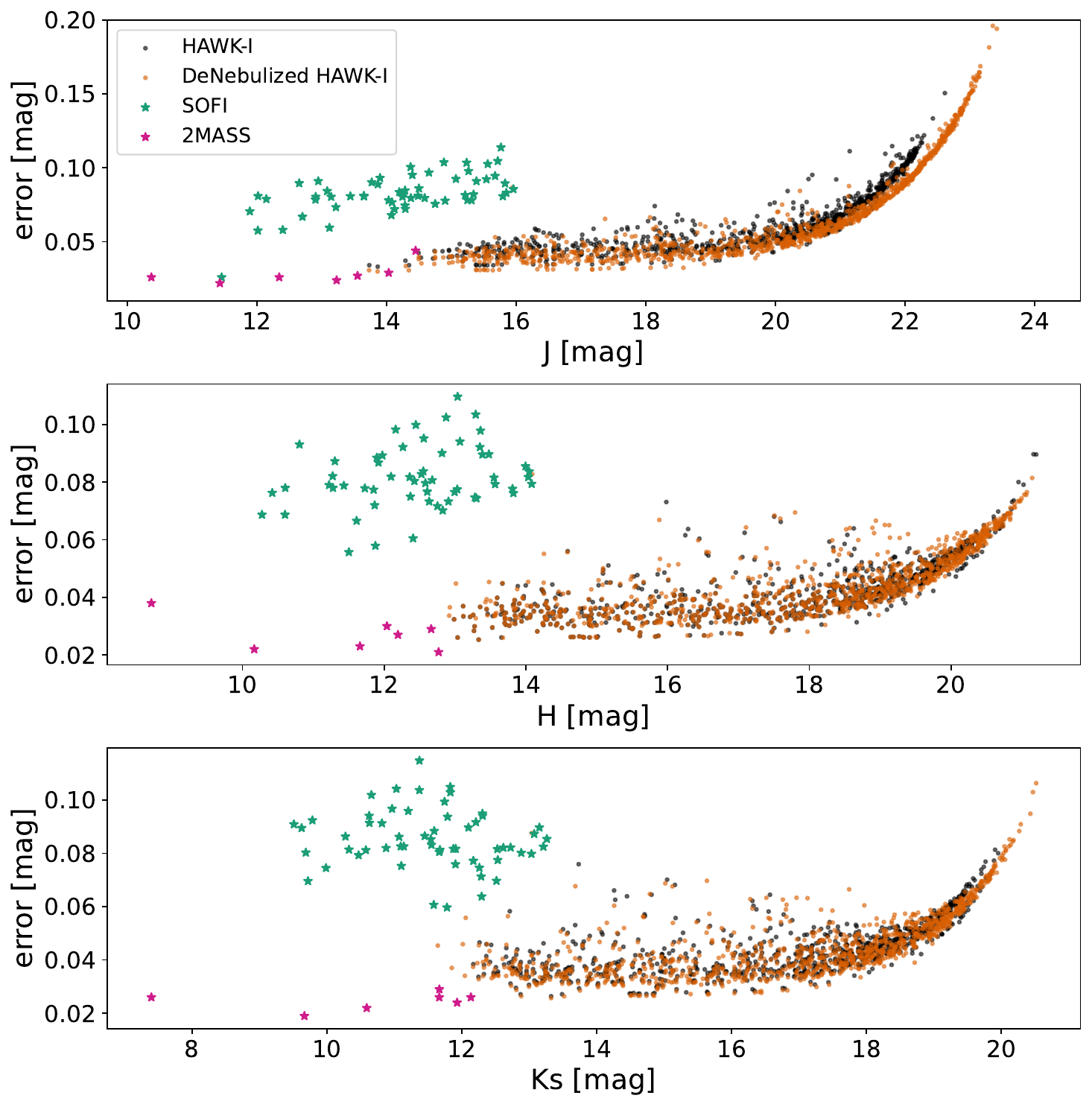}
    \caption{Photometric uncertainties associated with the magnitude data extracted from both the original HAWK-I data (in black) and the denebulised HAWK-I data (in orange) for the J-band (top), H band (centre) and $K_s$ band (bottom). Green stars are measurements from SOFI and purple stars are measurements from 2MASS.}
    \label{fig:pre_post_deneb_error}
\end{figure}

We extracted the photometry of the original HAWK-I data, the denebulised HAWK-I data and the control field data using the following photometric procedure. Point spread function (PSF) photometry was performed using {\sc SExtractor} \citep{sextractor}, and {\sc PSFEx} \citep{psfex}. The latter computes a spatially variable model  of the PSF, which is particularly important for datasets with complex PSF behaviours, such as the one presented here. A third-degree polynomial variation was used.
The {\sc SExtractor} output provides several parameters that are useful for quality control. We filter out all sources with keywords {\it FLAGS}$>3$ and {\it ELLIPTICITY}$\geq0.5$, which helps to remove galaxies, nebular clumps, knots and other spurious detections.\\
The photometric zero points and colour-terms were calculated from the comparison with the data from Two Micron All Sky Survey (2MASS; \citealt{2mass}) and a $JHK_S$ catalogue obtained from the SOFI  instrument \citep{sofi} at ESO's New Technology Telescope (NTT) described in Appendix~\ref{appendix:sofi}.
To avoid potential issues due to different spatial resolutions of the HAWK-I and calibration data, we discarded those sources with a neighbour at distances $<1.0''$ and brightness contrast $\Delta H<2$\,mag. Furthermore, we have rejected objects with uncertainties larger than 0.2 mag in the 2MASS and SOFI data, or in the HAWK-I instrumental magnitudes, as well as sources with 2MASS $ph\_qual$ flag different from A, B or C.\\
The photometric uncertainties were calculated by combining the uncertainties of the zero-points, colour-terms, and the measurement uncertainties supplied by {\sc SExtractor} (see Fig. \ref{fig:pre_post_deneb_error}). One of the HAWK-I chips has no overlap with the SOFI field, and therefore the number of stars available for calibration is significantly lower than for the other chips, resulting in larger photometric uncertainties. Some sources located towards the centre of HAWK-I chip 4 (centre of the cluster) are saturated, so we replace their photometry by the SOFI measurements when these are not saturated and 2MASS measurements when not. Finally, we keep only the sources located in regions of the maximum overlap of the dithered frames in each filter and exclude the region affected by the Moon-related artifact (white contours in Figure \ref{fig:rcw36_deneb}).\\
\subsection{Data improvement due to \textsc{DeNeb} }
Fig. \ref{fig:pre_post_deneb_error} shows how the photometric uncertainties change within the corresponding magnitude range for the $J-$, $H-$ and $K_s-$ bands in both the original and denebulised catalogues; we can see that the denebulised catalogue is deeper by about a magnitude and that, at least in the fainter end of the $J-$band, which is the band that the diffuse nebulosity affects the most, the uncertainties are slightly lower after applying \textsc{DeNeb}. However, because \textsc{DeNeb} allows for more faint sources to be detected and extracted by {\sc Source-Extractor} the median uncertainties for all bands are the same as in the original images: 0.06\,mag in $J-$, and 0.04\,mag in the $H-$ and $K-$bands.\\
For the original images, we report a total of 1302 detections in the $J-$band, 1932 in the $H-$band and 2023 in the $K_s-$band. For the denebulised images, we report a total of 1107 detections in the $J$-band, 1842 in the $H-$band and 1965 in the $K_s-$band. These differences are because of \textbf{false positives} in the original images from the presence of a bright nebula in our large 27.5 arcmin$^2$ on-sky area, and are expanded on in Appendix~\ref{appendix:Deneb_analysis}. After matching the different band catalogues within $1''$, the final $JHK_s$ catalogue from the original images contains 938 objects with data in all 3 bands, while the final $JHK_s$ catalogue from the denebulised images contains 1078 objects.\\
\subsection{Final $JHK_s$ catalogue}
We keep the $JHK_s$ denebulised data for all the analysis in this work. In addition to the 1078 sources detected in all three bands, there are 657 sources in the denebulised catalogue that match in the $H-$ and $K_s-$bands but are not detected in the $J-$band, which is the most affected by the high extinction in our field. The photometry of this $HK_s$ population is faint, with a mean $H-$band of 20.7 and mean $K_s-$band of 19.2, and less accurate, with a mean $H-$ and $K_s-$ band errors of $\sim0.13$; however, a visual inspection of the images revealed that they are real detections in both images, so we added them to our final catalogue. The final catalogue contains 1735 sources, of which 1078 have $JHK_s$ photometry.

\begin{table*}
\caption{Calibrated $JHK_s$ photometric catalogue of the sources in the HAWK-I field for the first ten point sources, sorting by right ascension.}
	\begin{center}
		\begin{tabular}{ccccccccc}
            \toprule
			$\alpha$ & $\delta$ & $J$ & $dJ$ & $H$ & $dH$ & $K_s$ & $dK_s$ & phot\_ref \\
            \addlinespace[5pt]
            \bottomrule
			08:59:11.69 & -43:41:44.4 & 20.36 &  0.05 & 19.52 &  0.04 & 18.98 &  0.05 & HAWK-I \\
			08:59:11.98 & -43:42:04.6 & 15.44 &  0.04 & 14.35 &  0.03 & 13.49 &  0.03 & HAWK-I \\
			08:59:12.34 & -43:42:53.0 & 12.40 &  0.06 & 11.88 &  0.06 & 11.78 &  0.06 & SOFI \\
			08:59:12.37 & -43:42:39.3 & 23.14 &  0.16 & 20.46 &  0.06 & 19.17 &  0.06 & HAWK-I \\
			08:59:12.52 & -43:41:36.0 & 15.72 &  0.04 & 14.45 &  0.03 & 13.73 &  0.03 & HAWK-I \\
			08:59:12.55 & -43:42:16.3 & 13.23 &  0.02 & 11.66 &  0.02 & 10.59 &  0.02 & 2MASS \\
			08:59:12.59 & -43:43:07.7 & 22.72 &  0.13 & 20.37 &  0.06 & 19.10 &  0.06 & HAWK-I \\
			08:59:12.60 & -43:42:01.3 & 16.57 &  0.03 & 15.87 &  0.03 & 15.67 &  0.03 & HAWK-I \\
			08:59:12.74 & -43:41:00.5 & 23.14 &  0.16 & 19.13 &  0.07 & 17.13 &  0.06 & HAWK-I \\
			08:59:12.84 & -43:43:34.9 & 15.50 &  0.03 & 14.75 &  0.03 & 14.56 &  0.03 & HAWK-I \\
		\end{tabular}
	\end{center}
    \tablefoot{The \textit{phot\_ref} column refers to the reference for the photometric data for that point source (see Section \ref{sec:photometry}). The full catalogue will be available digitally on CDS.}
	\label{tab:RCW36_JHKs_table}
\end{table*}

\subsection{Completeness}
\label{sec:completeness}
An artificial star test was used to assess the completeness levels of the photometry. 
Artificial stars were added to the original, denebulised and control field images using the open-source Python package Artificial Stellar Populations (ArtPop; \citealt{artpop}).
Photometry was then extracted from the generated image following the identical procedure, settings, and PSF solution as those used to obtain the photometry for the cluster and the control field (Section~\ref{sec:photometry}). We inserted 200 stars over the whole field, at a time, to avoid crowding and repeated the procedure 10 times at each magnitude (in steps of 0.2 mag) to improve statistics. \\
The completeness, defined as the ratio between the number of recovered and inserted artificial stars is shown in Appendix~\ref{fig:completeness} as a function of magnitude, for both the denebulised cluster (solid lines) and the original images (dashed lines). The 90\% and 50\% completeness limits for the original images, the denebulised images and the control field are given in Appendix~\ref{tab:completeness}.\\
We observed a decrease in the completeness of HAWK-I chip 4 (see Figures \ref{fig:onsky} and \ref{fig:rcw36_deneb}) when compared to the other chips, even in the denebulised images. In addition to being the chip in which the cluster is centred on, and thus the chip that presents higher extinction levels of the whole field, it is also the chip which presents more saturated pixels from the very bright pair of O9 stars present towards the cluster centre and other bright nebular features.\\
We also report that \textsc{DeNeb} is able to improve the completeness of each chip to the same level of the control field, except in chip 4. Possibly this is a consequence of the presence of saturation near the centre of the cluster itself and crowding instead of an inherent problem of \textsc{DeNeb} itself. In any case, we refer the reader to Appendix~\ref{appendix:Deneb_analysis} and Appendix~\ref{appendix_s:completeness} of the Appendix, which expands upon this discussion.
\section{Distance to RCW\,36 from Gaia DR3}
\label{s:gaia_distance}
To estimate the distance to RCW\,36, we queried the Gaia DR3 \citep{Gaia,Gaia_DR3} 
database within a radius of 10$'$ from the cluster position given in Table~\ref{tab:obs}. After filtering out the sources with the RUWE parameter > 1.4 (see \citealt{lindegren21}), we cross-matched this catalogue ($<1''$) with the lists of previously known member candidates from \citet{ellerbroek13} and \citet{broos13}, which resulted in a list of 117 unique member candidates. This list was then refined to include only candidates with the proper motions located within the 2$\sigma$ ellipse from the mean proper motion ($\mu_{\alpha}cos\delta$=$-6.9\pm 2.7$\,mas\,yr$^{-1}$, $\mu_{\delta}$=4.0$\pm$1.9\,mas\,yr$^{-1}$), and satisfying the criterion \begin{math}
\zeta = |\varpi-\overline\varpi|/\sqrt{\sigma_\varpi^2 + \sigma^2} >3
\end{math},
where $\varpi$ and $\sigma_\varpi$ represent the parallax measurement and uncertainty for each star, while $\overline\varpi$ and $\sigma$ are the weighted mean and standard deviation of the entire sample, correspondingly. This selection process is presented in Fig. \ref{fig:distance_gaiaDR3}.
Prior to applying the mentioned parallax criterion, we corrected the parallaxes for the bias described in \citet{lindegren21}, using the associated publicly available Python package\footnote{\url{https://gitlab.com/icc-ub/public/gaiadr3_zeropoint}}. 

\begin{figure*}
   \centering
   \center
\includegraphics[width=0.9\textwidth]{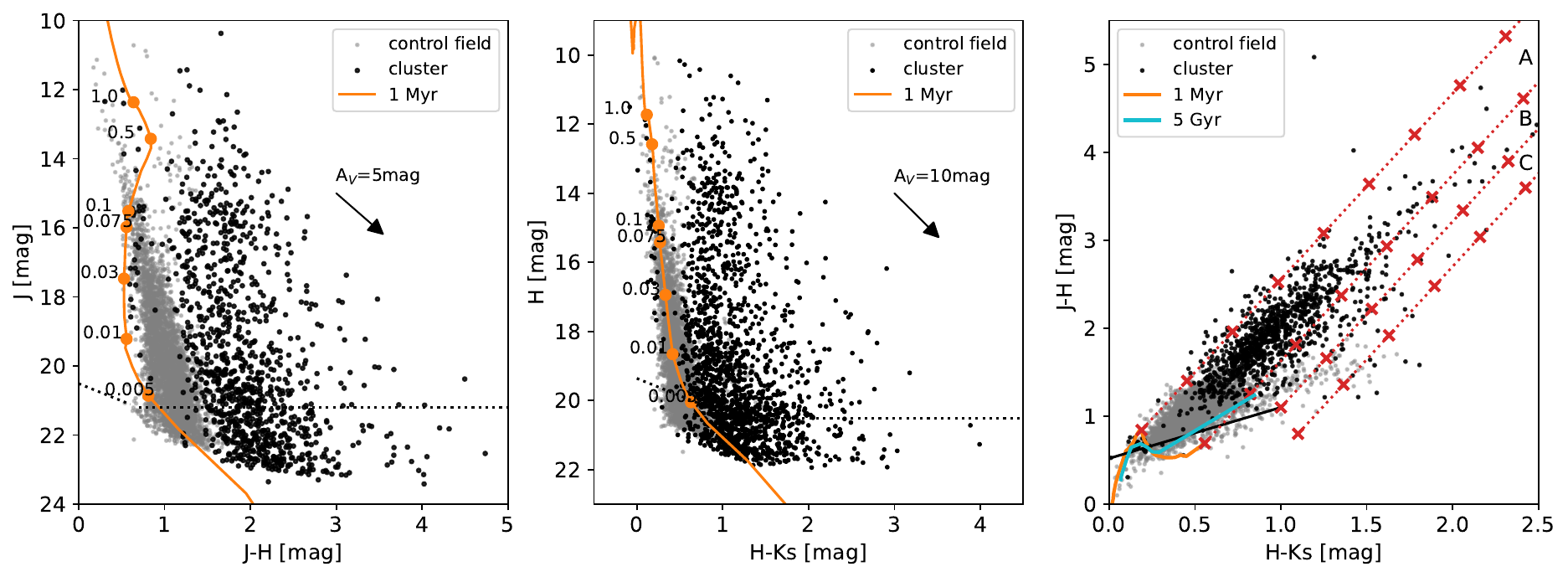}  
\hfill
\center
\includegraphics[width=0.9\textwidth]{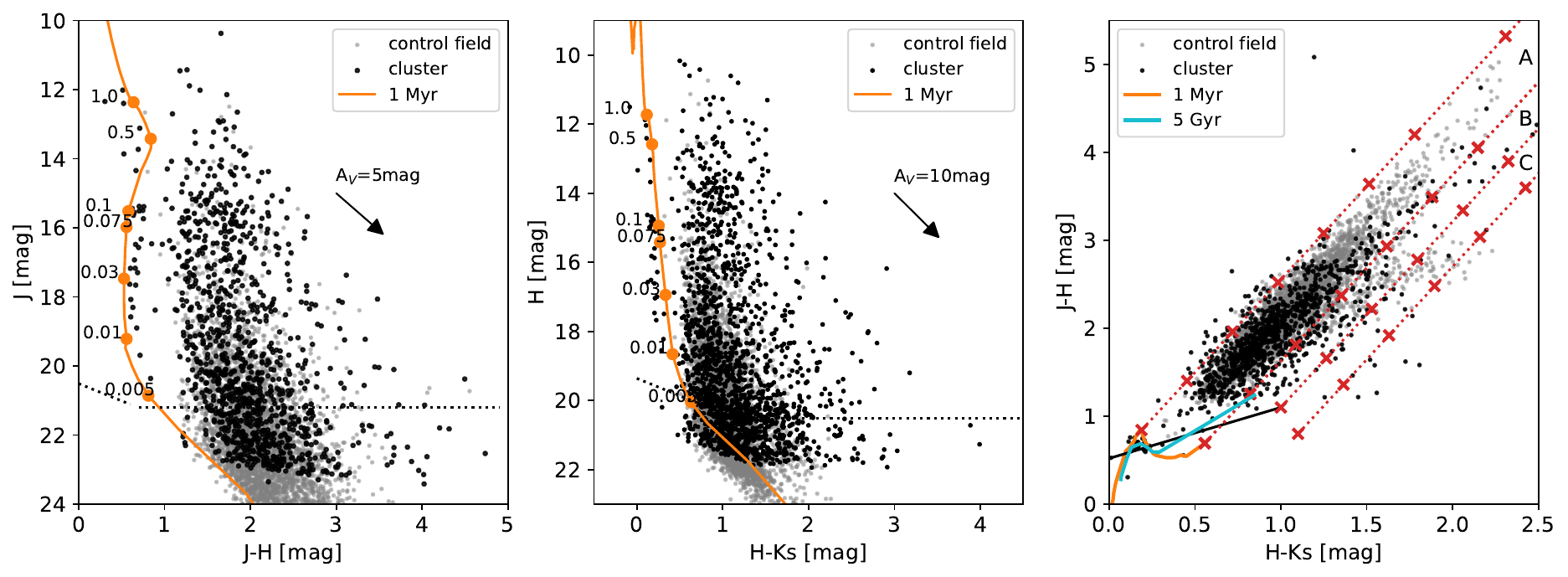}
\hfill

      \caption{Colour–magnitude diagrams of $J$, $J-H$ (left) and \textbf{$H$, $H- Ks$} (middle), and $J-H$, $H-Ks$ colour–colour diagram (right) of the sources detected toward RCW\,36 (black dots) and the control field (grey dots). The orange line shows the 1\,Myr isochrone shifted to the distance
of 950 pc, while the cyan solid line in the right-most panels represents the 5 Gyr isochrone used to estimate the extinction of the control field. The dotted red lines in the CCDs show the reddening vectors, and the red crosses along the reddening lines mark A$_V$ = 0 - 40\,mag, in step of 5\,mag. The black dotted lines mark the average 90\% completeness limit of the photometry. The solid black line represents the T Tauri loci from \cite{Meyer+1997}. The letter A refers to the region of giants and dwarfs, the letter B refers to the T Tauri region and finally the C region refers to the Herbig AeBe region \citep{Hernandez+2005}. The upper panels show original control field photometry, and the lower ones contain control field photometry reddened to match the distribution of the extinction in the cluster field (see Section \ref{sec:ext_diff}).
}
         \label{fig:cmds}
\end{figure*}

To estimate the distance, we employ the maximum-likelihood procedure as in \citet{Muzic+2019} and \citet{cantatgaudin18}, using the 88 probable members that pass the cuts described above. The obtained distance to RCW\,36 is $954 \pm 40$\,pc.\\

Accurate statistical distance determinations to RCW\,36 and other regions within the VMR are scarce. The region is sufficiently far and embedded so that not a lot of kinematical data exists, even in Gaia DR3. \cite{MurphyMay1991} were able to determine kinematical distance estimates using the relation between the rms line-of-sight velocity dispersion of the clouds and the rotation curve of the Milky Way; they inferred $\approx 1$ kpc to cloud C. \cite{Liseau92} performed a more thorough analysis and found a distance estimation of $\approx 0.7 \pm 0.2$ kpc for cloud C; this has been the canonical distance used as recent as \cite{ellerbroek13, Bordier+2024} and just barely agrees with our determination within the uncertainties. Possibly, an underestimation of the error in addition to the different methods employed could explain the discrepancy within both measurements.\\
In the last 20 years, there have also been a number of specific objects in cloud C to which the spectroscopic parallax has been measured. \cite{ellerbroek13} obtained X-SHOOTER spectra for the 4 sources with the brightest photospheric spectra in their catalogue, all corresponding to early type objects towards the centre of RCW\,36 and obtained distances from spectroscopic parallax ranging from $0.67$ kpc to $1.00$ kpc.\\
Our determination, which agrees with most of the literature, provides the most precise determination of the distance to RCW\,36, yet.

\begin{figure}
   \centering
   \includegraphics[width=0.45\textwidth]{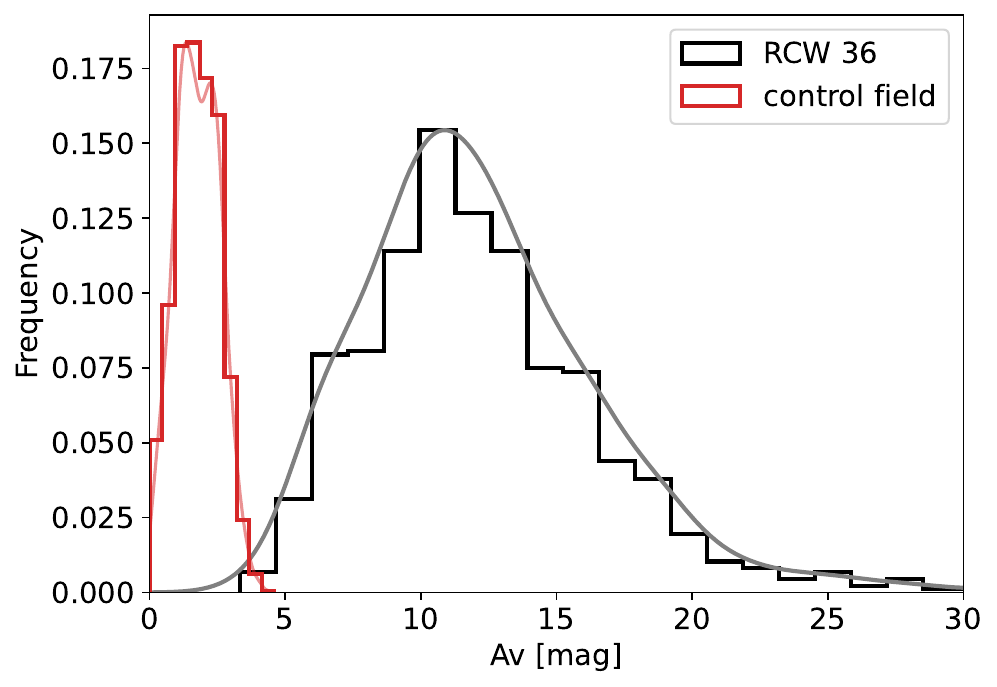}
      \caption{Distribution of the visual extinction towards RCW\,36 (black) and the control field (red). The smooth lines represent Gaussian KDEs of the two distributions.}
         \label{fig:extinction}
\end{figure}
\begin{figure*}
   \centering
   \includegraphics[width=\textwidth]{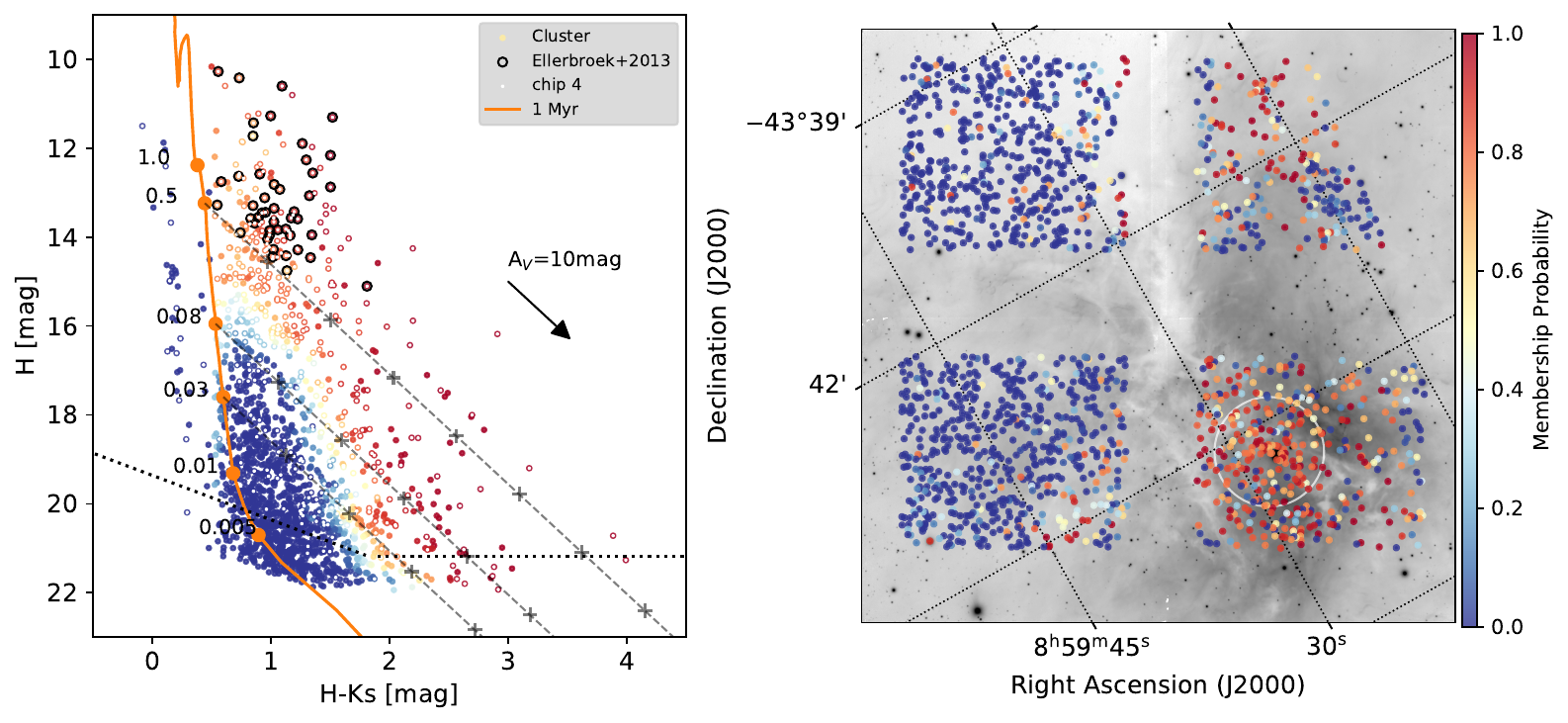}
      \caption{Colour-magnitude diagram (left) and the on-sky distribution (right) of the sources in our $JHK_s$ catalogue. The colour scheme represents the \textbf{membership weights computed in Section \ref{sec:memberhip_cmd} as a probability}. The orange solid line in the left panel is the 1\,Myr isochrone, the dotted line  refers to the 50$\%$ completeness limit for chip 4, and the dashed lines mark the position of a 0.5\,M$_{\odot}$, 0.08\,M$_{\odot}$ and 0.03\,M$_{\odot}$ source at different extinctions. The plus signs mark the A$_V$=[10, 20, 30, 40, 50, 60]\,mag. The black circles identify the 50 spectroscopically confirmed members from \cite{ellerbroek13}. The white dots identify the sources which are located within HAWK-I chip 4, in which the cluster is centred on. The white circle on the right is a visual representation of the inner region employed for the IMF shown in the bottom of Figure~\ref{fig:imf} and discussed in Section~\ref{ss:mass_segregation}.}
         \label{fig:probs}
\end{figure*}
\section{Membership}
\label{s:membership}
The sources belonging to RCW\,36 are mixed on the sky and in the colour-magnitude space with the field contaminants. In the CMD (Fig.~\ref{fig:cmds}, top left panel), a foreground sequence can be discerned on the blue side of the diagram and is easily corrected for, but separating cluster members from the background sources requires a more complex analysis as reddened background sources can mimic the colours and luminosities of genuine members. While, ideally, one would use spectroscopy for this purpose, it is in most cases extremely time-consuming and unfeasible in terms of source brightness at the faint end. We, therefore, employ a method in which the membership is determined statistically, through a comparison of the CMDs in the cluster direction with that in the direction of the control field.\\
However, before we can do that, we need to take into account potentially different amounts of extinction between the two fields, as the background stars and galaxies in the direction of the cluster will be reddened by a similar amount of extinction as the members. In Sec.~\ref{sec:ext_diff} we describe how we deal with the difference in extinction between the cluster and the control field, and in Sec.~\ref{sec:memberhip_cmd} we describe the method for statistical membership determination.

\subsection{Mimicking the cluster extinction distribution onto the control field}
\label{sec:ext_diff}
To estimate the extinction, we use only the sources with all 3 $JHK_s$ detections in the denebulised catalogue and de-redden \citep{Wang+2019} the photometry in the colour-colour diagrams (CCD) (Fig.~\ref{fig:cmds}, upper right panel) until it matches a model isochrone. We also only take into account for this procedure the sources that fall within the region which is described by the models. In the case of the cluster field, we employ an isochrone which has been created by combining different 1\,Myr isochrones for different mass ranges. These stellar evolutionary codes are PARSEC ($M>1M_{\odot}$; \citealt{Bressan+2012}), BT-Settl ($0.01M_{\odot}<M_*<1M_{\odot}$; \citealt{Baraffe+2015}) and AMES-Dusty ($M<0.01M_{\odot}$; \citealt{AMESDusty2001}). For the control field, we use a 5 Gyr model created analogously to the former one (cyan). \\
In Fig.~\ref{fig:extinction}, we show the A$_V$ distribution for the two fields, observing that the cluster distribution is more reddened and broader than the control field, with a difference of the medians of the two distributions of 7.8\,\textbf{mag}. The median $A_V$ values of the distributions for the RCW\,36 field and the control field are $\sim 11.7$\,\textbf{mag} and $\sim1.7$\,\textbf{mag}, respectively.\\

However, since the cluster field $A_V$ distribution is very broad we cannot simply redden the control field sources by a single value. Instead, we redden them in a way to approximate the cluster extinction distribution. This is done by randomly sampling from the cluster distribution represented as a Kernel Density Estimator (KDE) with a Silverman kernel \citep{Silverman1986}, and reddening each control field source by the difference of the obtained random value and its own extinction. For the sources for which we cannot obtain the extinction (located outside of the reddening band), we apply the mean difference of the two distributions. The result of this procedure can be seen in the lower panels of Fig.~\ref{fig:cmds}, where the grey dots represent the reddened control field sources.

\subsection{Membership determination from CMDs}
\label{sec:memberhip_cmd}

 For the statistical membership determination, we used a procedure identical to that described in \citet{Muzic+2019}. To summarise:
\begin{enumerate}
    \item A CMD is subdivided into grid cells with a step size ($\Delta$col, $\Delta$mag) in the colour and magnitude axes, respectively.
    \item The expected number densities of stars both in the cluster and the control fields is calculated for each CMD cell (as in \citealt{Bonatto_Bica2007}) and the number density of field stars is scaled to account for different on-sky areas covered by the images (due to the region excluded from the analysis described in Section \ref{sec:observations_and_data_reduction}) and the differences in completeness between the control field and the cluster field.
    \item A number of objects corresponding to the expected field population are then randomly removed from corresponding cells of the cluster CMD.

\end{enumerate}
We tested two different approaches of the method described above in different CMDs, both of which include a variation of the cell sizes ($\Delta$col = 0.3, 0.4, 0.5 mag and $\Delta$mag = 0.3, 0.4, 0.5 mag), a shift of the grid by $\pm$1/3 of the cell width in each dimension and 20 resamplings of the extinction correction described in \ref{sec:ext_diff}. \textbf{In the end, for each approach, we always arrive at 1620 lists of surviving members from which we compute the following statistics, the IMF and the star to brown dwarf ratio.}\\
The first method includes only the sources in the catalogue with photometry in $JHK_s$. Here, we decide to employ the J, J-H CMD. We estimate a median of $239^{+20}_{-16}$ surviving cluster members, where the uncertainty represents the 95\% confidence limits of the 1620 outcomes of the procedure.\\ 
The second method makes use of the $HK_s$ sources without a $J-$band detection. Here we perform the elimination procedure on the $H$, $H-K_s$ CMD. This results in a larger median number of surviving cluster members, $426^{+42}_{-65}$, when compared to the first method.\\
For each method, we estimate that about 77\% and 75\% of the sources observed toward the cluster are actually field contaminants, respectively. We find the overall ratio of members to be very similar between both methods (22\%-24\%), and decide to employ the $HK_s$ candidate members in the following analysis due to the higher number of members which enables our statistics to probe the lowest mass end of the IMF. \\
In Fig.~\ref{fig:probs} we show the CMD and the on-sky distribution of the sources in $HK_s$ catalogue, which were coloured according to \textbf{computed membership weights from the analysis above, as a probability}. The latter was calculated as the frequency at which each source appears in the 1620 lists. In the \textbf{right-hand} panel of Figure \ref{fig:probs}, the high-probability sources are strongly clustered in the lower right corner of the field (HAWK-I chip 4), while on the left side of the image (chips 1 and 2) we seem to observe mostly contaminants. More high probability sources are also located in the upper right corner of the field (HAWK-I chip 3), about $0.8\,$pc away from the cluster centre, following the filamentary nebula.
\subsection{Possible overestimation of the extinction field}
To investigate the possibility of overestimating the extinction field due to using a young isochrone on a field in which $\approx75\%-77\%$ of sources are older contaminants, we redo the same methodology described in Section \ref{sec:ext_diff} but using only the set of sources with membership \textbf{weights} larger than $50\%$. Although the distribution becomes less broad, the peak is still $A_V\approx$11\,\textbf{mag} and the median number of surviving sources in each member list remains the same after re-doing the procedure with the new distribution.\\
In fact, in Fig. \ref{fig:probs} we can see that the sources in HAWK-I chip 4 (white dots), where the cluster is centred, overlap quite well with the high probability sources and that their density in the plot decreases toward the low probability sources. Assuming most members are located close to the centre of the cluster, the sources located in chip 4 should map the region in the CMD where most members will be located, which is what we observe. This strengthens the claim that not only does our statistical elimination procedure work well in separating the members from the background contaminants but also that our extinction field distributions are close to reality.
\subsection{Comparison with former candidate members}
Lastly, by constraining the members from \cite{broos13} to the same area of our analysis (white contours in Figure \ref{fig:rcw36_deneb}) we find that we recover 152 sources within $1''$ from the 195 total in the area. From these 152 sources, only 10 sources have membership \textbf{weights} lower than 50\% and 79 sources have \textbf{weights} higher than 80\%. Of the 43 sources which do not match with \textbf{our} catalogue, we find that 15 are located near the edges of the regions of analysis and most other sources are located near the centre of the cluster, where our HAWK-I images present some saturation. We conclude that we recover most of the probable members from \cite{broos13}, which further \textbf{strengthens} the validity of our membership method.
\begin{table*}   
    \caption{Power-law slope $\alpha$ ($dN/dM \propto M^{-\alpha}$) of the IMF in various star forming regions along with characteristic parameters of each cluster.}
    \centering                                                                                                            
    \begin{tabular}{lccccc}                                                                                        
        \hline                                                                                                            
        Region & Distance [pc] & log($F_{FUV}$)$^f$& Surface density$^f$ [pc$^{-2}$] & $\alpha$ & Mass Range [M$_{\odot}$] \\ \hline
        \addlinespace[10pt]
        RCW\,36 & 954  & 4.2 & 451 & 1.62 $\pm$ 0.03 & 0.20 - 30 \\                                                         
               &      &      &      & 0.46 $\pm$ 0.14 & 0.03 - 0.20 \\                        \addlinespace[10pt]                                   
        RCW\,38 & 1700$^a$ & 5.4 & 3620 & 1.48 $\pm$ 0.08$^b$ & 0.20 - 20 \\                                                          
               &      &      &      & 0.51 $\pm$ 0.15$^f$ & 0.020 - 0.20 \\                       \addlinespace[10pt]                         
        ONC    & 400$^{c}$  & 3.3 & 350  & 2.3 $\pm$ 0.09$^d$ & 0.17 - 3.2 \\                                                           
               &      &      &      & 0.59 $\pm$ 0.06$^e$ & 0.0001 - 0.19 \\
                &      &      &      & 0.17 $\pm$ 0.04$^g$ & 0.003 - 0.20 \\
               
               \addlinespace[10pt]                           
        NGC\,1333 & 300 $^{c}$ & 1.4 & 200  & 0.48 $\pm$ 0.15 $^{f}$ & 0.010 - 0.20 \\                                                        
                  &      &      &  & 1.00 $\pm$ 0.1$^{f}$ & 0.10 - 1.0     \\ \hline                                                 
    \end{tabular}                                                 
    \\
    \tablebib{a) \cite{wolk08}; b) \cite{Muzic2017}; c) \cite{Kuhn+2019}; d) \cite{DaRio2012}; e) \cite{Gennaro2020}; ;f) \cite{Muzic+2025}; g) \cite{Strampelli2024}.}
    \label{tab:imf_slope_table}                                                             
\end{table*}
\section{Masses}
\label{ss:masses_ext}
For the mass determination we employed a Monte-Carlo procedure which is based on de-reddening photometry from the JHKs catalogue (1078 sources) to find the intersection of the data with the same 1\,Myr isochrone presented in Section \ref{sec:ext_diff} and the extinction law from \cite{Wang+2019}. Specifically:
\begin{enumerate}
    \item We checked if the data are optimal for the procedure by determining if the photometric data intersects with the model in the extinction vector direction, on the $J-H$, $H-Ks$ CCD. We also include sources which are located in the extended region of the T Tauri locus \citep{Meyer+1997} and the Herbig AeBe region from \cite{Hernandez+2005}. These regions are the A, B and C regions shown in the CCDs on the right, in Figure \ref{fig:cmds}.
    \item For each source, we sampled 1000 $J-$band and $H-$band values from a Gaussian distribution around the photometric points with the standard deviation corresponding to the photometric errors.
    \item In a $J$, $J-H$ CMD, we found the intersection of each point with the 1\,Myr isochrone, through the extinction vector direction. For each generated point we obtained a value of $A_V$ and the mass from the point where the photometry intersects with the isochrone.
\end{enumerate}
This procedure enables the possibility of determining a probability distribution function (PDF) of the mass and the extinction for each source. We find that the determined mass/extinction PDFs are not Gaussian and can sometimes have complex shapes with more than a single maxima. We thus keep the PDFs in order to sample masses for the determination of the IMF.\\
In order to get masses for the sources with only $HK_s$ photometry (657 sources), we also redo the same procedure in the $H$, $H-K_s$ CMD, without checking for an intersection with the isochrone on the $J-H$, $H-Ks$ CCD. In the analysis that follows, we always sample masses from the PDFs obtained using the $J$, $J-H$ CMD, except for the sources with no $J-$band detection, for which we employ the masses from the $H$, $H-Ks$ CMDs.\\

\begin{figure}
    \centering
    \includegraphics[width=0.45\textwidth]{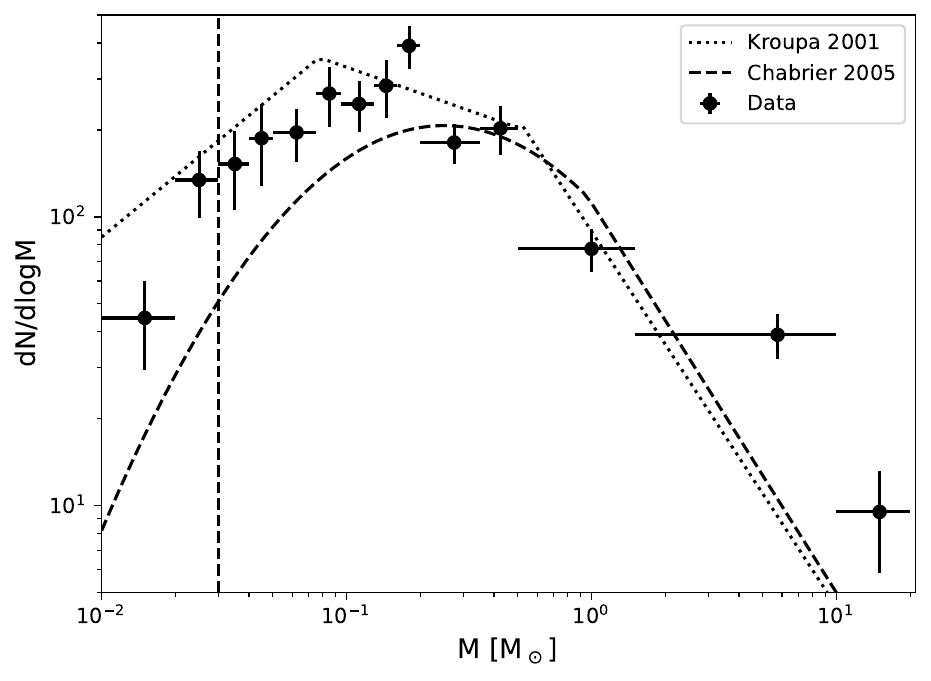}
    \includegraphics[width=0.45\textwidth]{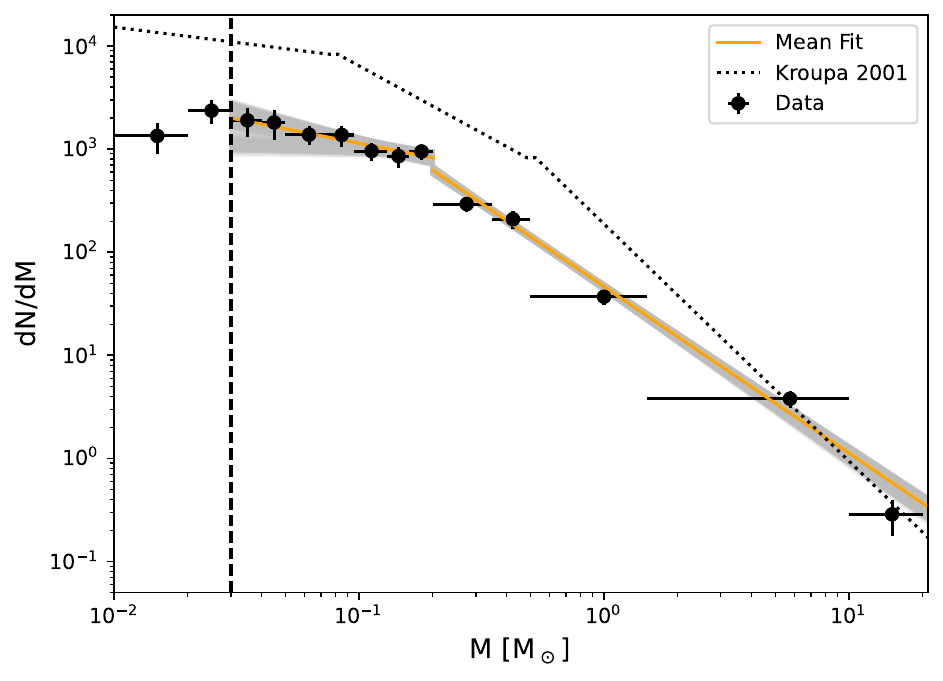}
    \includegraphics[width=0.45\textwidth]{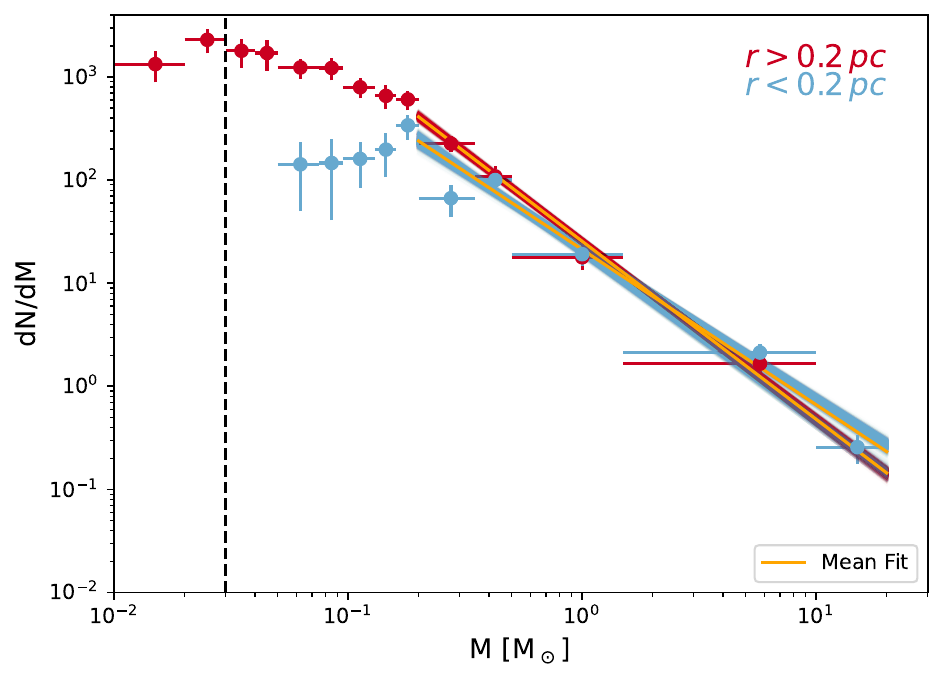}
    \caption{Completeness-corrected IMFs (up to 50\% limit) of RCW\,36. The vertical dashed line in all plots refers to the completeness limit assuming $A_V<12$\,\textbf{mag} (see discussion in Section \ref{ss:low_mass_imf}). The vertical solid error bars are the mean 1$\sigma$ computed for each bin, while the horizontal solid error bars refer to the bin widths. In the top plot we present the $dNdlogM$ form of the IMF in contrast to the \cite{Chabrier+2005} IMF (dashed black curve) and the \cite{Kroupa2001} IMF (dot-dashed curve). In the centre plot, we present the results from the broken power law fits described in Section \ref{s:imf}, with the orange line referring to the weighted mean $\alpha$ ($dNdM\propto M^{-\alpha}$) obtained from the distribution for all member lists. The distribution of slopes obtained is also shown as the grey area in the centre plot. In the bottom plot, we present the two IMFs obtained for the sources within (white circle in Figure~\ref{fig:probs}) and outside a $0.2$pc radius from the centre of the cluster. The red (outer region) and blue (inner region) areas in the bottom plot represent the whole distribution of the obtained $\alpha$. Bins that included no sources in at least one of the member lists were removed from the plots.}
    \label{fig:imf}
\end{figure}

\section{Cluster properties}
In order to disentangle possible differences in the shape of the IMF between different clusters it is first necessary to have a consistent and precise determination of astrophysical parameters that accurately reflect those differences. In addition, they should be determined using the same procedures as to avoid any biases which are inherent to different methods. Thus, to comment on possible differences due to the stellar density and the intensity of the radiation field we employ the same methodology described and employed in \cite{Muzic+2019, Muzic+2025} to determine the surface stellar density and the logarithmic far-ultraviolet flux ($\log$($F_{FUV}$)) for RCW\,36 and the ONC. These results are described in Appendix~\ref{appendix:surface_density} and Appendix~\ref{appendix:far_uv_flux} and shown in Table \ref{tab:imf_slope_table} \textbf{in addition to the values for NGC\,1333 and RCW\,38 from \cite{Muzic+2025}}.

\section{Initial mass function}\label{s:imf}
For each list of surviving sources, we sampled 100 masses for each source from the mass distributions. Each of the resulting 100 lists of possible masses is then also bootstrapped with repetition 100 more times. For each of the 1620 lists of surviving members, we arrive at 10 000 different IMFs. We make use of the $H-$ band completeness function (see Appendix~\ref{fig:completeness}) to correct for the chip dependent completeness. This means that, for a source with apparent magnitude corresponding to the 50\% $H-$ band completeness, we would add two sources in the same corresponding IMF mass bin, instead of one. We perform this correction up to the $H-$band 50\% completeness limit in each chip (see Appendix~\ref{tab:completeness}), eliminating sources below these limits from the IMF analysis. Also, because the pair of central O9 stars \citep{Bik2005, ellerbroek13} are not present in our catalogue (see discussion in Appendix~\ref{appendix_s:completeness}), we manually add both stars to the highest mass bin ($10M_{\odot}-20M_{\odot}$) of each IMF\footnote{Using the O star parameters as a function of the spectral type presented in \cite{Martins+2005} in addition to the spectral types from \cite{Bik2005, ellerbroek13} we can estimate these objects to have a mass of $\sim18M_{\odot}$ each.}\\
We also select the bins of the IMF so that each bin has the same average number of sources ($\sim25$ sources per bin; see the following works for a more in depth discussion of how equally spaced binning can bias the determination of the $\alpha$, \citealt{Offner2014, Maiz2005}, and \cite{Genaro+2021} for a discussion on a possible method to correct for this bias when using equally spaced bins). \\
The completeness-corrected IMF is shown in Figure \ref{fig:imf} in both $dN/dM$ (centre) and $dN/d\log M$ (top) forms. \\
To determine the $\alpha$ ($dN/dM \propto M^{-\alpha}$), we calculate the average of the 10 000 values for each bin and then determine the slope for each of the 1620 IMFs. This allows us to get a final distribution of slopes and errors from which we can determine the weighted mean value and the weighted standard deviation. The results from these fits are presented in Table \ref{tab:imf_slope_table} and showcased in the centre plot of Figure \ref{fig:imf}, with the orange line referring to the weighted mean $\alpha$ fit for the distribution of 1620 member list IMFs (see Section \ref{sec:memberhip_cmd}) while the grey density area maps the whole range of slopes from the obtained distribution. We report that our IMF is well represented by a broken power law with 0.2 $M_{\odot}$ as a break point, and we find a significant difference of over $5\sigma$ between the two slopes and a flattening for the lower masses close to the expected values from the literature (see Figure 1 of \citealt{Hennebelle+2024} and references therein, in addition to our Table \ref{tab:imf_slope_table}).

\subsection{\textbf{High-mass} slope}
When computing $\alpha$ within $0.2M_{\odot}-20M_{\odot}$ we find that increasing the minimum mass in the fit range always yields the same results, with $\alpha$ never as steep as the \cite{Salpeter1955} slope ($\alpha=2.35$). Such shallow slopes for the higher mass regime of the IMF are, however, not uncommon in YMC. In the core of RCW\,38, a much denser and bright young cluster than RCW\,36, also located in the VMR, the $\alpha$ found in the mass range $0.20M_{\odot}-20M_{\odot}$ \citep{Muzic2017} is similar to what we find in the same mass range (see Table~\ref{tab:imf_slope_table}). Other similar shallow slopes can also be found in distant YMC such as Westerlund I, another young massive cluster in the Milky Way located at a distance of $4\,$kpc ($\alpha\approx1.8$ within $5M_{\odot}-100M_{\odot}$; \citealt{Lim2013}), the Arches cluster, $8\,$kpc towards the \textbf{Galactic} centre ($\alpha \sim 1.8$, for the mass range $1.8M_{\odot}-51M_{\odot}$; \citealt{Hosek2019}) and in Trumpler\,14 ($\alpha \sim 1.7$, for the mass range $0.2M_{\odot}-4.5M_{\odot}$; \citealt{Rom+2025}).\\
Nearby ($<500\,$pc), we can also find such shallow slopes in the Orion complex, in the low intensity extinction field cluster ($A_V\lessapprox1$\,\textbf{mag}), Collinder\,69, located in $\lambda\,$Orionis ($1.7\lessapprox\alpha\lessapprox1.9$; \citealt{Bayo+2011}), and in $\sigma\,$Orionis ($1.6\lessapprox\alpha\lessapprox1.9$; \citealt{penaramirez12}).
In the ONC, which has a comparable surface density but a magnitude lower $F_{FUV}$ to RCW\,36, both \cite{Muench2002} and \cite{DaRio2012} find a \textbf{high-mass} slope which is comparable to Salpeter. In the same cluster, however, \cite{Gennaro2020} find shallower slopes to be possible in a Markov chain Monte Carlo analysis in which they introduce and fit for both the binary fraction, the nuisance parameters and different star formation history cases. Another case of Salpeter like \textbf{high-mass} slope can also be found in NGC\,2244 ($1.5M_{\odot}-20M_{\odot}$; \citealt{Muzic2022}), although albeit quite bright, this region has a much lower surface density than the previous examples \citep{Muzic+2025}.
\subsection{Low-mass slope}
\label{ss:low_mass_imf}
We computed the $\alpha$ within $0.03M_{\odot}$-$0.2M_{\odot}$ to understand if we also observe a flattening ($0<\alpha<1$) when compared the \textbf{intermediate to high-mass} regime. We chose the minimum mass as $0.03M_{\odot}$ because it represents the 50\% completeness limit of the H-band, transformed into mass via the isochrones described in Section~\ref{sec:ext_diff}, assuming $A_V\sim12$\,\textbf{mag}. In fact, even with a completeness correction up to the 50\% limits of the $H-$band, the variability of the extinction field also causes our completeness to vary. For sources with $A_V\lessapprox12$\,\textbf{mag}, we should be complete down to $0.03M_{\odot}$; however, for the most reddened sources ($A_V\gtrapprox12$\,\textbf{mag}), our completeness decreases to $M\sim0.15M_{\odot}$ for $A_V=30$\,\textbf{mag}.\\
We observe a flattening in line with what is found in other regions, regardless of their inherent properties. In YMC such as RCW\,38 ($0.36\lessapprox\alpha\lessapprox0.66$ in the range $0.02M_{\odot}$-$0.20M_{\odot}$; \citealt{Muzic+2025}), Collinder\,69 ($0.2\lessapprox\alpha\lessapprox0.4$ in the range $0.01M_{\odot}$-$0.65M_{\odot}$; \citealt{bayo11}), the ONC ($0.1\lessapprox\alpha\lessapprox0.7$; \citealt{Gennaro2020, DaRio2012}) and Trumpler\,14 ($0.1\lessapprox\alpha\lessapprox0.5$ in the range $0.03M_{\odot}$-$0.2M_{\odot}$; \citealt{Rom+2025}) this behaviour is observed with values similar to the ones obtained in this work. In clusters with lower surface density and $F_{FUV}$ than RCW\,36, the same flattening of the slope is also observed. In NGC\,1333, \cite{Muzic+2025,scholz12a} find a similar, but even shallower $\alpha$ below 0.2$M_{\odot}$ ($\alpha=0.48\pm0.15$) and 1$M_{\odot}$ ($0.61 \pm 0.08$). The same behaviour is also observed in Cha-I and Lupus\,3, \citep{Muzic+2025, muzic15} and also Corona Australis \citep{Muzic+2025, Muzic2022}, which all show a slope under 0.2$M_{\odot}$ that is $0<\alpha<1$.\\

\section{Star to brown dwarf ratio}
\label{ss:star_bds_ratio}
We find the lowest mass member of our catalogue \textbf{with a high membership weight (90\%)} to have a median mass of $\approx0.036$ $M_{\odot}$; well within the brown dwarf mass regime (i.e $M<0.075M_{\odot}$). In total we find 19 sources with membership \textbf{weights} higher than 90\% which have median masses below the hydrogen-burning limit.\\
In order to estimate the star to brown dwarf ratio (star-BD ratio) \textbf{we employ the 1620 lists of members from Section \ref{s:membership}} in a Monte Carlo procedure like in Section \ref{s:imf} \textbf{(resampling 100 masses, which are bootstrapped 100 times)}, then identify the stars ($0.075\leq M\leq1$) and brown dwarfs ($0.02-0.03\leq M\leq0.075$), \textbf{calculate the mean star-BD ratios for each list of members} and determine the medians and 95\% confidence intervals; these values are shown in Table \ref{table:cluster_properties}. \\
In this analysis we have also varied the minimum mass of brown dwarfs from $0.03M_{\odot}$ to $0.02M_{\odot}$ to study how this change would impact our analysis. We find that this mass significantly impacts the determinations of the number of brown dwarfs and thus the star-BD ratio; however, both median values are located within the 95\% confidence limits of each other. Overall we find that the star-BD ratio for RCW\,36 should be between $\sim2-5$ as it is also found in regions with high surface density and radiation field ($2.1 \pm 0.6$ for RCW\,38; \citealt{Muzic+2025,Muzic2017}) and their less dense and radiative counterparts ($3.2 - 4.8$ for Cha-I and $2.1 - 4.5$ for Lupus\,3; \citealt{muzic15,Muzic+2025}). Our median value is, nevertheless, slightly larger, possibly, because: we are overestimating the contaminants and thus losing possible brown dwarf candidates; the spatially variable extinction field can cause the completeness to also change spatially; the uncertainties in the models and photometry (due to these sources having a higher photometric error due to being fainter). These same problems also explain the high confidence limits for the ratio and the number of stars and brown dwarfs as is discussed in the literature (\citealt{Muzic+2025} and references therein)
\begin{table}                                                                          
\caption{Variation of the number of stars, BDs, and the star-BD ratio with the minimum BD mass.}                                                           
\label{table:cluster_properties}                                                       
\centering                                                                             
\begin{tabular}{lcc}                                                                   
    \hline\hline  
    \addlinespace[2pt]
    Stars ($M_{\odot}$) & $0.075<M<1.0$ & $0.075<M<1.0$ \\
    \addlinespace[5pt]
    BDs ($M_{\odot}$) & $M>0.03$ & $M>0.02$ \\                       
    \hline           
    \addlinespace[10pt]
    Number of stars & $227^{+17}_{-17}$ & $227^{+17}_{-17}$ \\        
    \addlinespace[10pt]    
    Number of BDs & $73^{+15}_{-30}$ & $98^{+22}_{-45}$ \\
    \addlinespace[10pt]
    Star-BD ratio & $3.2^{+2.1}_{-0.5}$ & $2.3^{+1.9}_{-0.4}$ \\
    \addlinespace[10pt]
    \hline                                                                      
\end{tabular}                                                                        
\end{table}
\section{Mass segregation}
\label{ss:mass_segregation}
Kinetically evolved clusters should present a higher density of \textbf{high-mass} stars on the central part of the cluster than on the outskirts. This property is called mass segregation. On the other hand, YMC might also present a higher density of massive stars in the centre since their inception. This property is called primordial mass segregation (see review by \citealt{Portegies+2010}).\\
To assess the possibility of mass segregation in RCW\,36 we calculate the $\alpha$ using the same methodology employed and described in Section \ref{s:imf} for the sources within a radius of 0.2\,pc of the centre of the cluster\footnote{The centre of the cluster was selected visually from the region of higher saturation in the centre of the cluster.} and outside of it. Figure~\ref{fig:probs} showcases the inner region of the analysis in the white circle on the plot to the right and the choice of the radius is discussed in Appendix~\ref{appendix:mass_segregation}. The two IMFs are shown in the bottom plot of Figure \ref{fig:imf}. The IMF slopes were obtained in the mass range $0.2M_{\odot}-10M_{\odot}$, to be sure that we are complete even up to $A_V~30$\,\textbf{mag} (see Appendix~\ref{appendix:mass_segregation} and the discussion in Section~\ref{ss:low_mass_imf}), and are 1.50 $\pm$ 0.07 for the sources in the inner $0.2$\,pc region of RCW\,36 and 1.73 $\pm$ 0.02 for the sources in the outer region. The obtained values of $\alpha$ do not agree within the uncertainties with the inner region presenting a shallower slope when compared to the outer region.\\
As another test, we employ the same methodology as \cite{Muzic+2019} and determine the cumulative distributions of mass for \textbf{low-} and \textbf{high-mass} stars up to $1\,$pc away from the same centre of the cluster employed previously. To do this, we employ the set of sources with membership \textbf{weight} probabilities higher than 80\% and calculate the Andersen-Darling statistic for both distributions to determine the p-value for the hypothesis that the distributions originate from the same underlying distribution. The results from this analysis are shown in Appendix~\ref{fig:mass_segregation_cumulative} and yielded small p-values of 0.01\footnote{Testing for more restrictive sets of input sources ($90\%$) or different break masses (1.0 $M_{\odot}$, 0.5 $M_{\odot}$) yields the same p-value.}.\\
Our results strengthen the claims by \cite{Baba04} of RCW\,36 having primordial mass segregation due to its young age $\leq1.1$\,Myr \citep{ellerbroek13,Kuhn+2015III, Sano2018}. However, caution must be taken because both our results and the findings in \cite{Baba04} could be biased towards massive stars near the centre of the cluster. In our case, because of saturation (see discussion in Appendix~\ref{appendix_s:completeness}) and \textbf{crowding toward the cluster centre. In the case of \cite{Baba04}, due to same bias of massive stars in the cluster centre, worse crowding and also a possible overestimation of the completeness}. In any case, the shallower than Salpeter \textbf{high-mass} $\alpha$ for the whole HAWK-I field, the different $\alpha$ determined in the inner and outer regions of the cluster and the cumulative distributions of high and low mass sources all point to possible mass segregation in RCW\,36.
\section{Summary and conclusions}
\label{s:conclusions}
Our results can be summarized as follows:
\begin{itemize}
    \item We employed a novel deep learning algorithm \textsc{DeNeb} to improve the extraction of the photometry on new GLAO observations of RCW\,36 using HAWK-I and obtain a $HK_s$ catalogue of 1735 sources, of which 1078 also have $J-$band detections.
    \item  We determined the distance to RCW\,36 using the previous known members from \cite{broos13, ellerbroek13} and kinematics from Gaia DR3. Our result, $954 \pm 40$ pc is the most precise measurement of the distance to RCW\,36 yet.
    \item Using lists of possible statistical members, we found a rich population of stars and brown dwarfs that extends more than a parsec away from the centre of the cluster.
    \item We determined the first IMF for RCW\,36 and found the $\alpha$ in the \textbf{high-mass} regime shallower than the Salpeter value but in agreement with many observed slopes in other YMC. In the lower mass regime we also observe the flattening of the slope which is found in other clusters and associations, regardless of their surface density and/or $F_{FUV}$.
    \item We determined statistics for the number of stars and brown dwarfs in the cluster. From this information we calculated the star-BD ratio to be 2-5, which is in agreement with the ranges found for other clusters.
    \item We found evidence of possible mass segregation in RCW\,36 from different $\alpha$ for the outer ($>0.2$\,pc) and inner regions ($<0.2$\,pc) of the cluster and from the different cumulative mass distributions of high- and low-mass sources.
\end{itemize}

\begin{acknowledgements}
This work has made use of data from the European Space Agency (ESA) mission
{\it Gaia} (\url{https://www.cosmos.esa.int/gaia}), processed by the {\it Gaia}
Data Processing and Analysis Consortium (DPAC,
\url{https://www.cosmos.esa.int/web/gaia/dpac/consortium}). Funding for the DPAC
has been provided by national institutions, in particular the institutions
participating in the {\it Gaia} Multilateral Agreement.\\
This work was supported by Fundação para a Ciência e a Tecnologia (FCT) through the research grants doi.org/10.54499/2022.03809.CEECIND/CP1722/CT0001, https:/doi.org/10.54499/2023.01915.BD and UID/04434/2023
V.A-A acknowledges support from the INAF grant 1.05.12.05.03\\
RS acknowledges  financial support from the Severo Ochoa grant CEX2021-001131-S funded by 658 MCIN/AEI/ 10.13039/501100011033 and from grant PID2022-136640NB-C21 funded by MCIN/AEI 10.13039/501100011033 and by the European Union\\
A.B. acknowledges support from the Deutsche Forschungsgemeinschaft (DFG, German Research Foundation) under Germany's Excellence Strategy - EXC 2094 - 390783311 and the Swedish National Space Agency (grant 2022-00154)\\
L.A.C. acknowledges support from ANID, FONDECYT Regular grant number 1241056 and the Millennium Nucleus on Young Exoplanets and their Moons (YEMS), ANID -NCN2024 001\\
\end{acknowledgements}

\bibliographystyle{aa} 
\bibliography{rcw36}
\clearpage
\appendix
\section{SOFI photometry}
\label{appendix:sofi}
Near-infrared images in $J-$, $H-$ and $K_S$-band have been downloaded from the ESO archive (programme ID 080.C-0836). We reduced the data applying the standard infrared data reduction procedures including cross-talk correction, flat-fielding, sky subtraction, bad pixel correction and stacking into a single deep mosaic per filter. The field covered in these observations spans $\sim 7'\times 7'$, centred at the cluster, only partially overlapping with the Hawk-I data.
The {\sc DaophotII/Allstar} standalone PSF-fitting algorithm \citep{stetson87} was used to extract the photometry, with a Gaussian PSF which varies quadratically with the position in the frame ($VARIABLE$ $PSF$ parameter set to 2). We filtered out all the sources with the sharpness parameter $|sh|>0.7$, which helps to remove galaxies, clumps, knots, and spurious detections caused by ghosts.
Photometric zero-point and colour-terms have been calculated through comparison with the 2MASS data. We have excluded objects with uncertainties larger than 0.2 mag in the 2MASS or in the SOFI instrumental magnitudes, as well as sources with 2MASS $ph\_qual$ flag different from A, B or C. The limiting magnitudes of the SOFI catalogue are $J\sim$20.5, H$\sim$18.5 and K$_S\sim$18\,mag, i.e. about 2 magnitudes shallower from the HAWK-I catalogue.
\section{HAWK-I data before and after \textsc{DeNeb}}
\label{appendix:Deneb_analysis}
\begin{figure*}[!h]
    \centering
    \includegraphics[width=\linewidth]{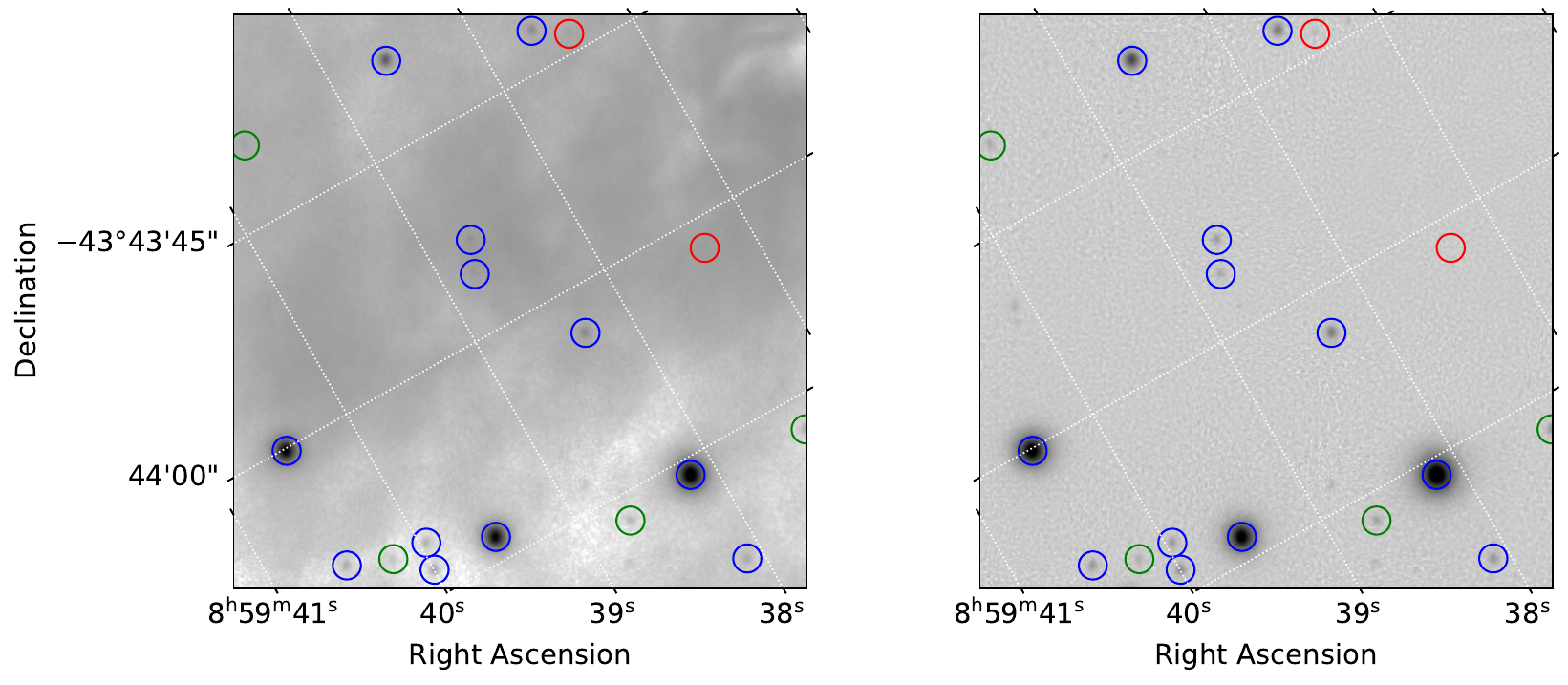}
    \caption{Cutouts of the original (left) and denebulised (right) HAWK-I $J-$band images. Circles in blue refer to the sources which are present in both catalogues. Red circles refer to the sources that are only present in the original $JHK_s$ catalogue. Green circles refer to the sources only present in the denebulised $JHK_s$ catalogue.}
    \label{fig:deneb_appear_disappear_sources}
\end{figure*}
In order to inspect the differences between the original and denebulised catalogs, we matched all sources with detections in all 3 bands within $1''$. We find that \textsc{DeNeb} enables the detection of 202 new sources not present in the catalogue extracted from the original $JHK_s$ images. These newly detected sources have a median J-band magnitude of 22.6 mag with a corresponding median uncertainty of 0.11 mag. Conversely, 62 sources present in the original data are no longer detected after \textsc{DeNeb} processing. These removed sources have a median J-band magnitude of 21.2 mag with a median uncertainty of 0.07 mag and are primarily associated with erroneous extractions linked to features of the bright nebula, such as knots; however, we also found some sources that don not appear in the denebulised data due to a bad extraction. This could be due to a noisier background around it or due to it having a bad or too complex PSF profile. Regardless, the amount of sources which are able to be extracted after applying \textsc{DeNeb} rather outweighs the much lower number of sources that suffer a bad extraction after \textsc{DeNeb} and not in the original images.\\

Figure~\ref{fig:deneb_appear_disappear_sources} shows two cutouts of the same region in both the original and denebulised images. The green circles, which correspond to the sources only detected in the denebulised images, are faintly observed in contrast with the nebula around them in the original data but in the denebulised images, without the nebula, they are clearly observed, and thus {\sc Source-Extractor} is now able to identify and extract their photometry. The red circles, which correspond to the sources that are identified and extracted in the original images but not on the denebulised images, are mostly not observed in contrast with the nebula in the original images and very clearly seem to be associated with nebula features such as knots, and in the denebulised images we can clearly see that, without the background, those positions are not related to any point sources whatsoever.\\
\begin{figure}
    \centering
    \includegraphics[width=0.45\textwidth]{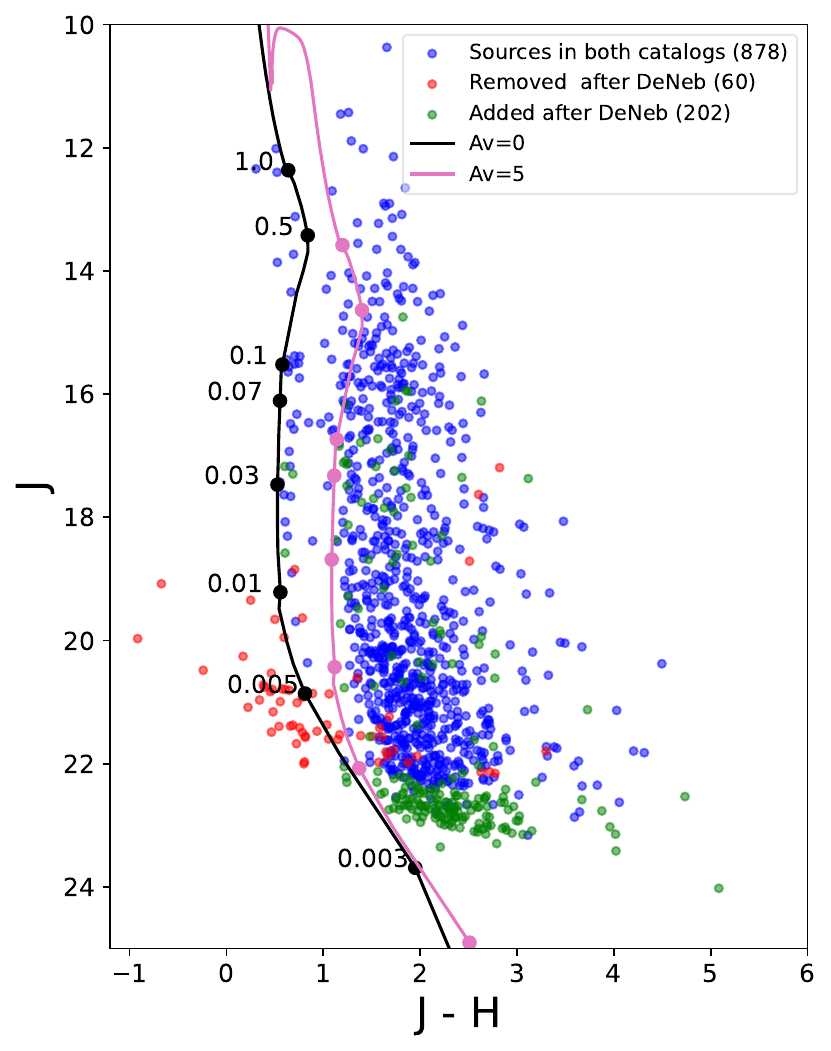}
    \caption{$J$, $J-H$ CMD showcasing the photometry of the \textbf{different populations of sources explained in Appendix \ref{appendix:Deneb_analysis} and showcased in Figure \ref{fig:deneb_appear_disappear_sources}}. The red points refer back to the sources identified only in the original images, while the green points refer back to the sources only identified in the denebulised images. The blue points are the sources which appear on both the original and denebulised images. The black curve is the unreddened 1 Myr isochrone plotted for a distance of 950 pc, while the magenta curve is the same isochrone plotted for the same distance and $A_V=5$\, \textbf{mag}.}
    \label{fig:cmd_appear_disappear_sources}
\end{figure}
In addition to visually inspecting the differences between the images, we have also studied the photometry of these different populations. In Fig. \ref{fig:cmd_appear_disappear_sources} we show the $J$, $J-H$ CMD for the same sources as before, with the same colour scheme. The sources that are removed after employing \textsc{DeNeb} have much bluer colours than the rest of the sources in the catalogs, with some having almost null or even negative colours, which is a behaviour that is contained to this population specifically (most other sources have a $J-H$ colour close to unity or larger, except for the clear foreground population which can be observed in the CMDs in Figure \ref{fig:cmds}). The sources which are identified only in the denebulised images, on the other hand, seem to follow the shape of the isochrone, are mostly located towards the fainter end of the CMD and show redder colours similar to the sources that appear in both the original and the denebulised catalogues. 

\section{Completeness of the HAWK-I data}
\label{appendix_s:completeness}
\begin{figure*}
   \centering
   \includegraphics[width=0.75\textwidth]{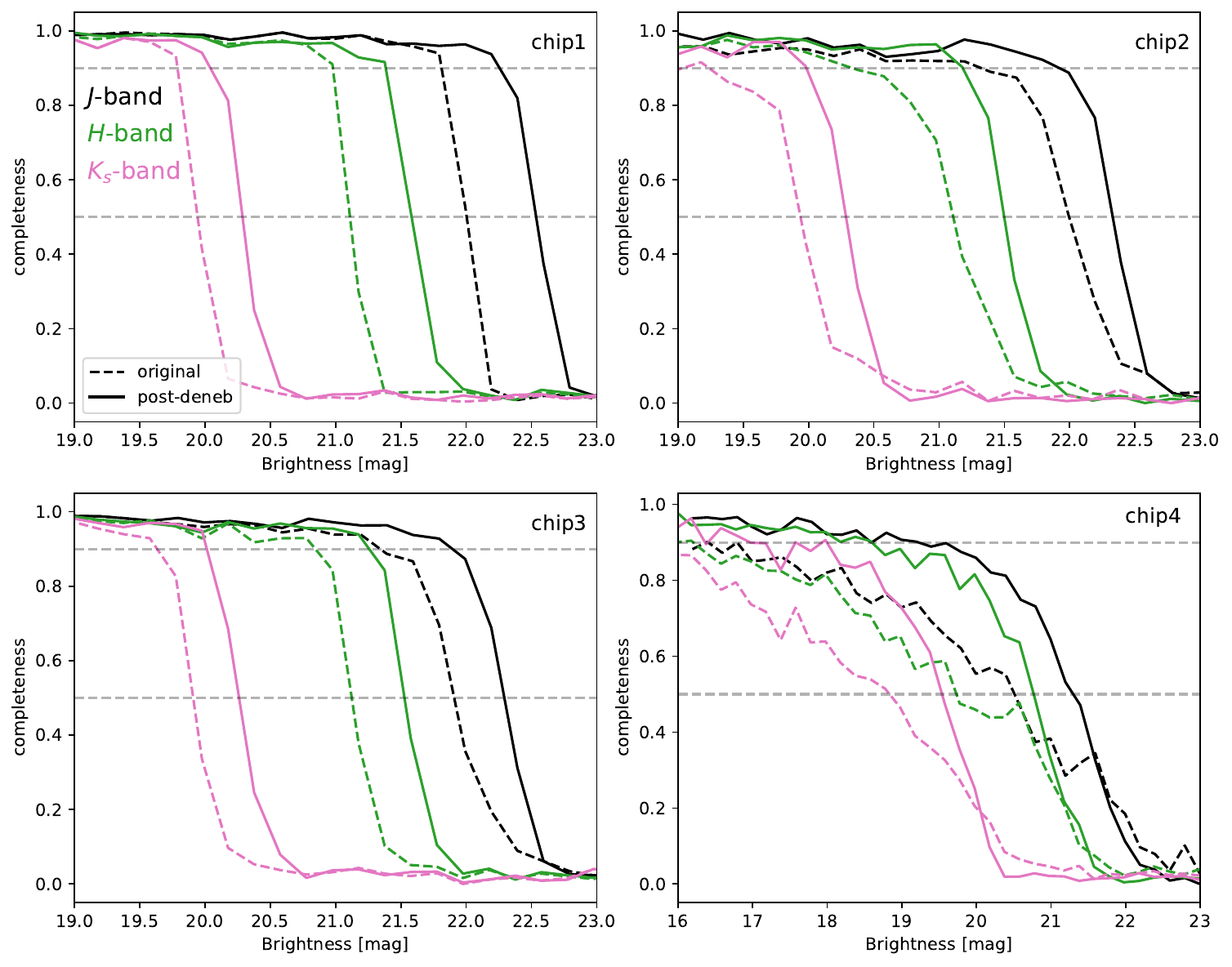}
      \caption{Chip dependent completeness plots as a function of apparent $JHK_s$ magnitudes in the original HAWK-I images and the denebulised HAWK-I images. Completeness was determined through the generation of 50 artificial sources in each of the HAWK-I $J-$,$H-$ and $K_s$- band image chips and our ability to recover these sources using the same photometric procedure employed and described in Section \ref{sec:photometry}. }
         \label{fig:completeness}
   \end{figure*}

The procedure to calculate the completeness limits for the HAWK-I data are described in Section~\ref{sec:completeness}. The completeness curves for each chip and filter are shown in Fig.~\ref{fig:completeness} for the original HAWK-I images and the denebulised HAWK-I images. The corresponding 90 and 50 per cent limits in addition to the limits for the control field (see Section~\ref{sec:observations_and_data_reduction}) are given in Table~\ref{tab:completeness}. We find that \textsc{DeNeb} allows for the improvement of depth over all four chips. However, only in chip 4 does this improvement not match or surpass the completeness limits of the control field, which is not denebulised.\\
In HAWK-I chip 4, which is the chip that the cluster is centered on, we find a 0.7 magnitude deeper 50\% completeness limit when comparing the denebulised images with the original ones, in the J band. Nevertheless it is still about a magnitude less deep than the control field in the same chip. This is most likely due to the fact that this is the only chip that presents saturated features in the chip. These are located in the location of the bright late O stars and also in a nearby bright nebular feature. Both of these saturated regions in the CCD cannot be fixed through the application of \textsc{DeNeb}, which means that, in our artificial star tests, any artificial source that is created randomly in these regions will not be recovered at all. This is true both in the original images and the denebulised images, however, and not a consequence of applying \textsc{DeNeb}. \\
In the first case, the artificial sources will be put into a region in which most nearby pixels will be very bright or saturated, and thus even if {\sc SExtractor} does identify (erronously) and extract data for a source, it will be discarded.\\
In the second case, \textsc{DeNeb} will simply remove all of the very bright or saturated pixels and replace them with random noise. Thus, also, any source that is in that region would also be removed and not extracted. \\
It can also be seen in Figure \ref{fig:probs} of the main body that there are regions in HAWK-I chip 4 in which no sources are detected, regardless of their membership \textbf{weights}.These regions trace the bright filamentary nebula in the original images. The same behaviour can also be seen in HAWK-I chip 3. However, this chip does not present nebular features as bright as the ones seen in HAWK-I chip 4 and specially no saturation at all. This explains the lower completeness limits of HAWK-I chip 4 in comparison with the remaining chips, even in the denebulised data.\\
To overcome this, we employ the SOFI catalogue described in \ref{appendix:sofi} for the sources which are saturated in the HAWK-I images and also 2MASS when the same sources are also saturated in the SOFI data. By comparing the final catalogue with the known members from \cite{ellerbroek13} we find that we are only still missing 3 objects: the pair of late O9 stars and a G star, all located in the region of saturation in the HAWK-I images.
   
\begin{table}[]
\centering
\caption{Completeness levels of the photometry.}
\begin{center}
\begin{tabular}{cccc|ccc}\hline\hline
chip  & J & H & K$_S$& J & H & K$_S$\\
\hline
\addlinespace[2pt]
 & \multicolumn{6}{c}{Original HAWK-I} \\
 & \multicolumn{3}{c}{90$\%$} &\multicolumn{3}{c}{50$\%$} \\
\hline
 1 & 22.2 & 21.2 & 19.8 & 22.4 & 21.4 & 20.0\\
 2 & 21.5 & 20.8 & 19.6 & 22.2 & 21.1 & 19.9  \\
 3 & 21.6 & 20.4 & 19.4 & 22.2 & 21.1 &  19.9 \\
 4 & 18.1 & 17.2 & 16.6  & 21.1 & 20.1 & 19.2\\
 \hline
 \addlinespace[2pt]
 & \multicolumn{6}{c}{Denebulised HAWK-I} \\
 & \multicolumn{3}{c}{90$\%$} &\multicolumn{3}{c}{50$\%$} \\
\hline
 1 & 22.3 & 21.4 & 20.0 & 22.5 & 21.6 & 20.3\\
 2 & 21.9 & 21.2 & 20.0 & 22.3 & 21.5 & 20.3  \\
 3 & 21.9 & 21.3 & 20.0 & 22.3 & 21.5 &  20.3 \\
 4 & 18.7 & 18.2 & 17.4  & 21.3 & 20.8 & 19.5\\
\hline
\addlinespace[2pt]
 & \multicolumn{6}{c}{Control field } \\
 & \multicolumn{3}{c}{90$\%$} & \multicolumn{3}{c}{50$\%$}  \\
 \hline
 1 & 21.6 & 20.7 & 19.9 & 22.2 & 21.2 & 20.1 \\
 2 & 21.8 & 20.9 & 19.8 & 22.1 & 21.1 & 20.1 \\
 3 & 21.7 & 20.9 & 20.0 & 22.1 & 21.1 & 20.2 \\
 4 & 21.4  & 20.8 & 19.9 & 22.2 & 21.2 & 20.1 \\
\hline
\end{tabular}
\end{center}
 \label{tab:completeness}
\end{table}
\section{Distance}
\begin{figure*}
    \centering
    \includegraphics[width=0.45\linewidth]{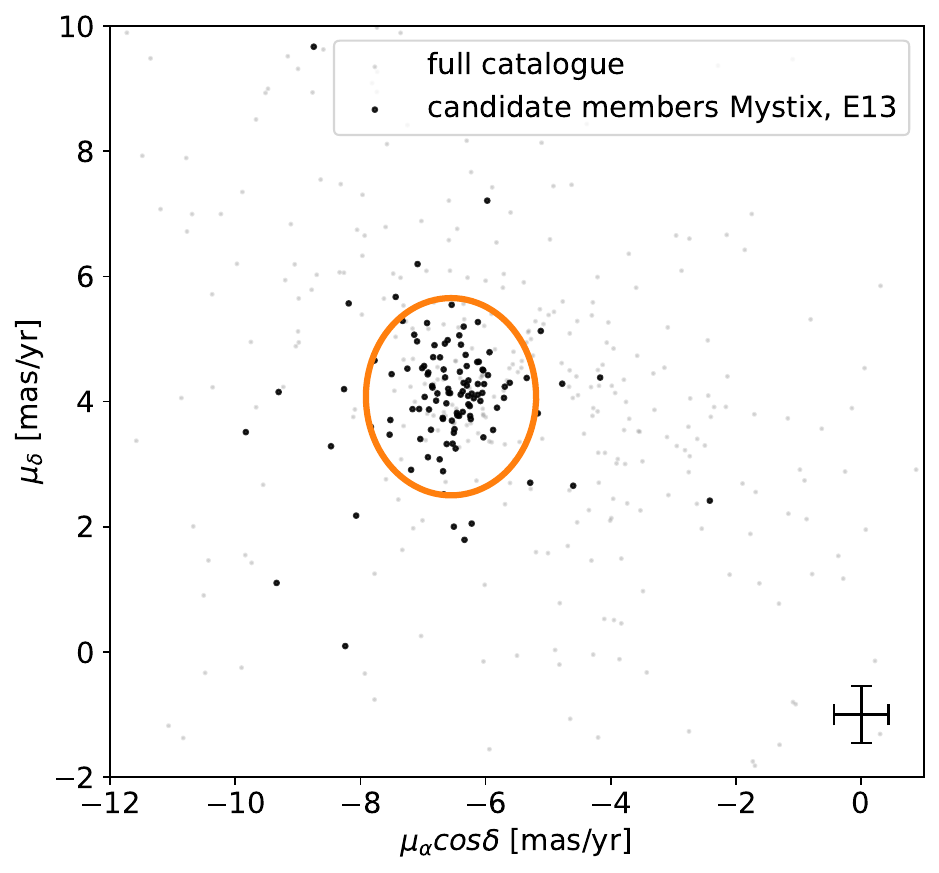}
    \includegraphics[width=0.45\linewidth]{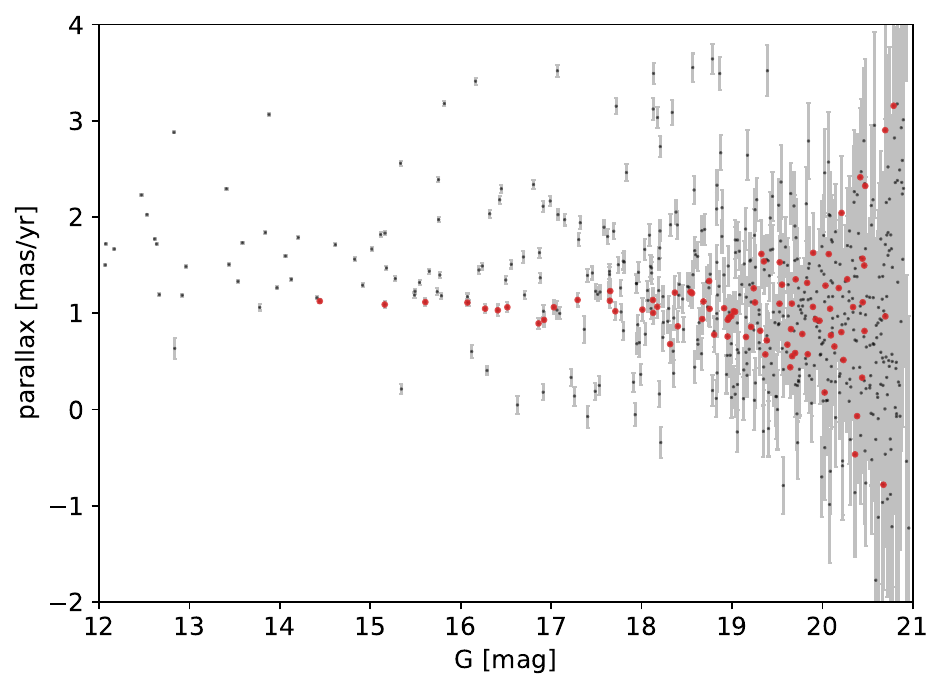}
    \caption{On the left, the ellipse marks the proper motion selection criteria for the probable members (black dots) of RCW\,36 from \cite{ellerbroek13, broos13}; the grey dots are the sources from Gaia DR3 \citep{Gaia, Gaia_DR3} in the  10$'$-radius field around the central position of RCW\,36. On the right, the sources selected based on the parallax and proper motion criteria (red dots; see Section \ref{s:gaia_distance}); black dots with grey error-bars represent the parallax and the corresponding uncertainties of the sources in the  10$'$ field.}
    \label{fig:distance_gaiaDR3}
\end{figure*}

In Figure \ref{fig:distance_gaiaDR3}, on the left, we show a visualization of the proper motion criteria while on the right we present the Gaia DR3 parallaxes of the sources surviving both the proper motion and parallax criteria, described in Section \ref{s:gaia_distance}.\\
\textbf{The choice of ellipse axes was intended to prioritize the purity of the sample used for estimating the cluster distance, rather than completeness. This criterion does not imply that all sources outside the ellipse are unrelated to the cluster. Approximately 80\% of these outliers are at the faint end of the Gaia catalogue (G > 19), where astrometric uncertainties are significant. A smaller fraction, about 5\%, shows highly discrepant parallaxes and proper motions, which may indicate association with a different foreground young population.}

\section{Surface density}
\label{appendix:surface_density}
\begin{figure*}
    \centering
    \includegraphics[width=\linewidth]{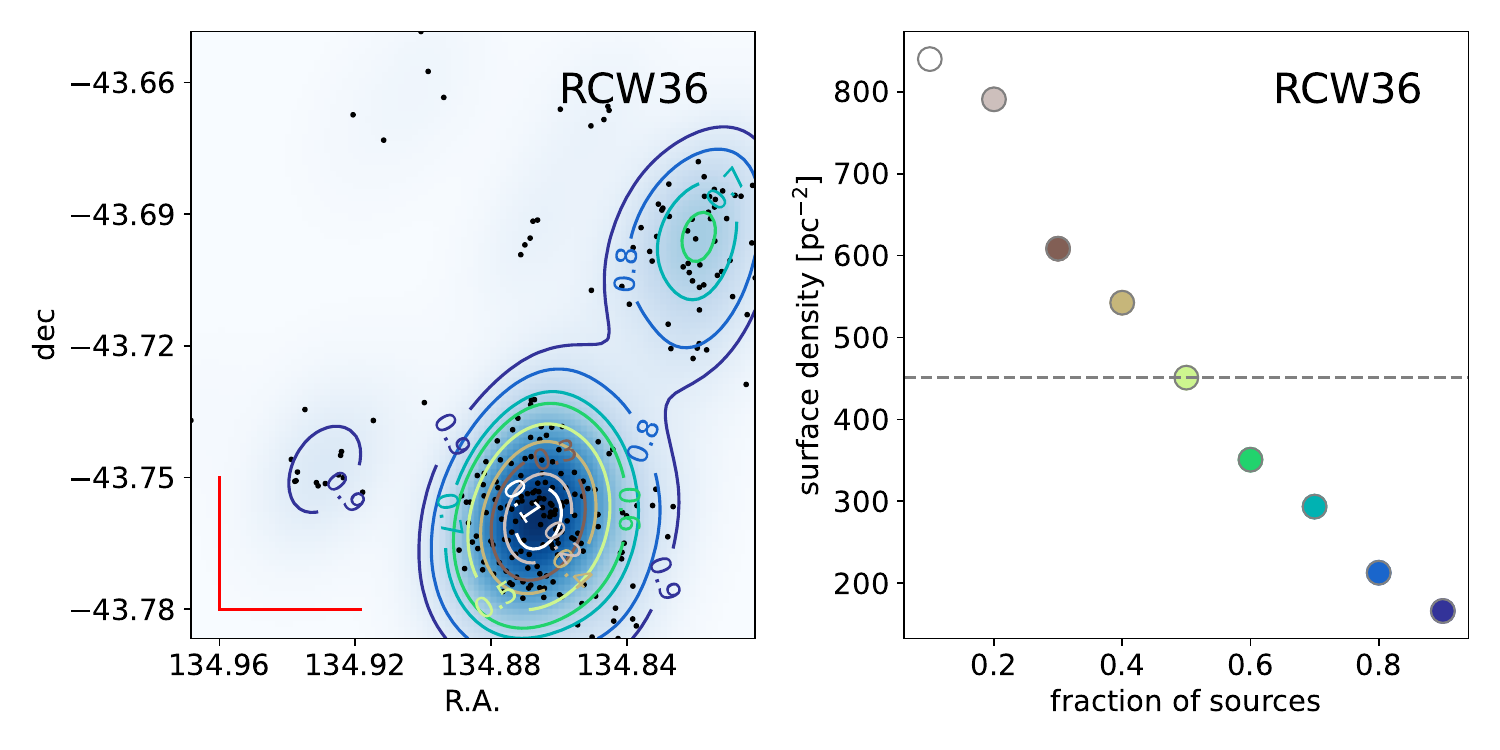}
    \caption{Spatial distribution for the set of sources in our HAWK-I data with membership \textbf{weight} probabilities higher than $80\%$. The color maps are two-dimensional KDEs of the candidate member distributions, and the contours represent the levels containing different percentages of the sources (10\%
to 90\%, in steps of 10\%). The stellar surface density for the area contained inside each contour are shown to the right of the KDE plot. The surface density associated with the 50\% contour are marked by the horizontal dashed lines.
The red bars shown in the lower right corner represent the 0.5 pc scale at the distance of RCW\,36.}
    \label{fig:surface_density}
\end{figure*}

The surface density for RCW\,36 was determined by employing two-dimensional KDEs of the sources in our catalogue with \textbf{membership weight} probabilities higher than 80\%; Figure \ref{fig:surface_density} showcases the surface densities associated with different level contours containing increasing fractions of the total number of sources. The surface density obtained for both clusters at the $50\%$ contour level is shown in Table \ref{tab:imf_slope_table} together with values from \cite{Muzic+2025} for other clusters.\\
We report that RCW\,36 has a surface density of 451 sources per $pc^{-2}$ at the 50\% contour level. However a slightly lower inner core density reaching 800 sources per parsec squared. These values put RCW\,36 at the medium to high end of values found for other clusters in \cite{Muzic+2025} with only RCW\,38 having a higher surface density, albeit a magnitude larger one. In the same magnitude order of RCW\,36 we find the ONC, which is about 25\% lower, and NGC\,1333, which is 77\% lower.
\section{Far-ultraviolet flux}
\label{appendix:far_uv_flux}
The far-ultraviolet fluxes in Table \ref{tab:imf_slope_table} for RCW\,36 and ONC were determined \textbf{using the list of members from \cite{DaRio2012} and the sources from our catalogue with membership weight probabilities larger than 80\%, together with OB star information for each cluster (ONC: \citealt{Quintana+2025}, RCW 36: \citealt{Bik2005})}. The log($F_{FUV}$) are shown for both clusters in Table \ref{tab:imf_slope_table} together with values from \cite{Muzic+2025} of other clusters.\\ 
RCW\,36 is found to have a log($F_{FUV}$) of 4.2, a higher value than both the ONC (3.3) and NGC\,1333 (1.4), which have $\sim1$ (ONC) to $~3$ (NGC\,1333) orders of magnitude lower log($F_{FUV}$). Although the latter clusters have a higher amount of OB stars in general, those stars are also more sparsely distributed. RCW\,36, on the other hand, is much smaller than both clusters, which means its FUV flux is more localized and intense.
\section{Mass segregation}
\label{appendix:mass_segregation}
\begin{figure*}
    \centering
    \includegraphics[width=0.45\textwidth]{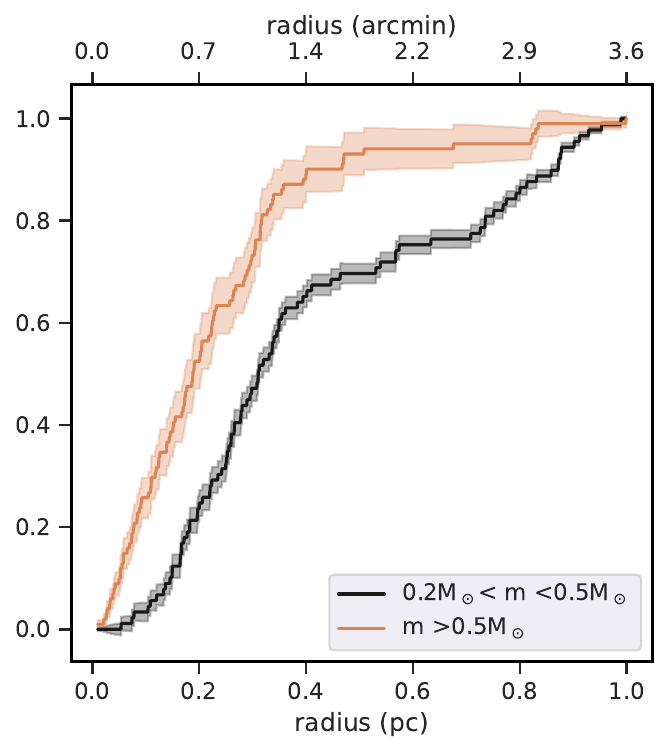}
    \includegraphics[width=0.45\linewidth]{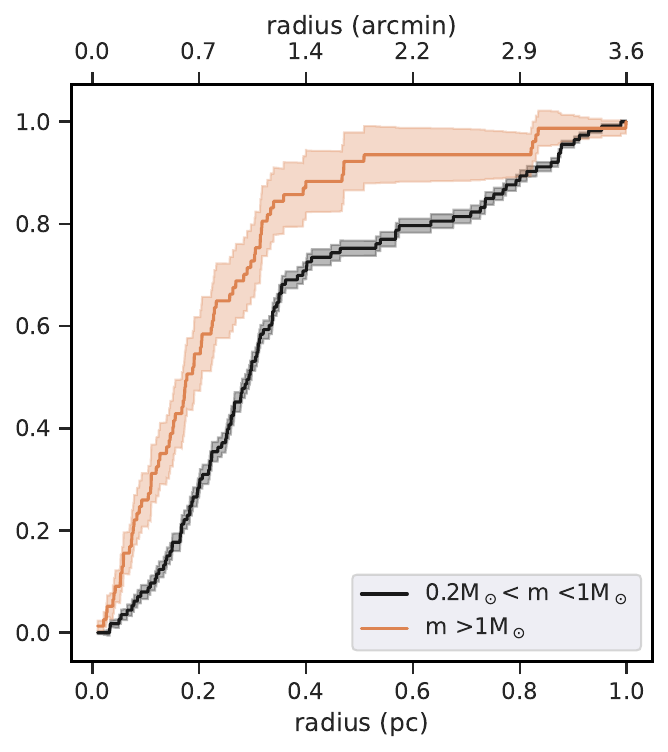}
    \caption{Cumulative mass functions for the set of sources in our HAWK-I data with membership \textbf{weights} larger than $80\%$ corresponding probability. Orange and grey regions trace the variability from 100 different mass sampled from the PDFs presented and described in Section \ref{ss:masses_ext} of the main body, which are bootstrapped with repetition another 100 times. {In dark orange and black the cumulative curves calculated from the median masses are shown for the different mass regimes.} On the left, we test the amount of sources with masses between $0.20M_{\odot}<M<0.5M_{\odot}$ against sources with masses higher than $0.5M_{\odot}$. On the right, the same test is redone for masses between $0.20M_{\odot}<M<1.0M_{\odot}$ and masses higher than $1.0M_{\odot}$.}
    \label{fig:mass_segregation_cumulative}
\end{figure*}

To investigate the possibility of mass segregation in RCW\,36 we performed two different analysis, both of which are described and discussed in depth in Section \ref{ss:mass_segregation}. The first is to determine the IMF for a set of sources in the inner part of the cluster and the outer part of the cluster. \textbf{We chose the inner and outer threshold to be 0.2\,pc because this allows for similar fluctuations of the amount of sources to populate the chosen mass regime for the inner and outer IMF while also not increasing the region too much so that it would comprise most of the known cluster. We report that changing the radius 0.3\,pc or 0.15\,pc yields consistent results to the one presented in Section \ref{ss:mass_segregation}. The white circle in the right-hand panel of Figure~\ref{fig:probs} represents the inner 0.2\,pc region.} In this analysis we employ a different mass regime\textbf{, compared to the one employed in Section~\ref{s:imf},} in which we are certain to be complete even for very high $A_V\sim30$\,\textbf{mag} \textbf{because we want to determine if there is a clear difference in the distribution of lower and intermediate-high mass stars in our field}. We thus compute the $\alpha$ within $0.2M_{\odot}<M<20M_{\odot}$, using the same lists of members and mass distributions employed in Section \ref{s:imf}. These results are shown in the bottom plot of Figure \ref{fig:imf}.\\
The second is to determine the cumulative mass distributions for the low and \textbf{intermediate to high-mass} sources. Similarly to the former procedure, in order to constraint the analysis to a mass range in which we know we are complete, we decide to define low mass as sources with mass within $0.2M_{\odot}<M<M_{break}$ and \textbf{intermediate to high-mass} as sources with mass $M_{break}<M<20M_{\odot}$. As to also study the possibility of $M_{break}$ changing our results significantly, we decide to test a number of different break masses in this analysis ($M_{break}[M_{\odot}]={0.2,0.5,1,2}$), all of which yielded high Andersen-Darling statistics which then resulted in low p values of $\sim0.001$. Figure \ref{fig:mass_segregation_cumulative} shows these results for $M_{break}[M_{\odot}]=\{0.5,1\}$.
\end{document}